\documentclass[reprint, amsmath,amssymb, prb,superscriptaddress]{revtex4-2}

\usepackage{graphicx}
\usepackage{dcolumn}
\usepackage{bm}
\usepackage[usenames,dvipsnames]{xcolor}

\PassOptionsToPackage{hyphens}{url}\usepackage{hyperref}
\hypersetup{
	colorlinks=true,
	linkcolor=RoyalBlue,
	filecolor=RoyalBlue,      
	urlcolor=RoyalBlue,
	citecolor=RoyalBlue
}

\usepackage[inline]{enumitem}
\usepackage{bbold}
\usepackage{soul}
\usepackage{placeins}
\usepackage{amsmath}
\usepackage{amssymb}
\usepackage{physics}
\usepackage{placeins}
\usepackage[normalem]{ulem}

\newcommand{\up}{\uparrow}
\newcommand{\upup}{{\uparrow\uparrow}}
\newcommand{\dn}{\downarrow}
\newcommand{\dndn}{{\downarrow\downarrow}}
\newcommand{\updn}{{\uparrow\downarrow}}
\newcommand{\dnup}{{\downarrow\uparrow}}

\begin{document}

\title{From perfect to imperfect poor man's Majoranas in minimal Kitaev chains}

\author{Melina Luethi}

\author{Henry F. Legg}
\author{Daniel Loss}
\author{Jelena Klinovaja}

\affiliation{Department of Physics, University of Basel, Klingelbergstrasse 82, CH-4056 Basel, Switzerland}

\date{\today}

\begin{abstract}
Poor man's Majoranas (PMMs) hold the promise to engineer Majorana bound states in a highly tunable setup consisting of a chain of quantum dots that are connected via superconductors. Due to recent progress in controlling the amplitudes of elastic cotunneling (ECT) and crossed Andreev reflection (CAR), two vital ingredients for PMMs, experimental investigations of PMMs have gained significant interest. Previously, analytic conditions for the ``sweet spots'' that result in PMMs have focused on systems with infinite Zeeman energy. Here, we derive analytically a sweet spot condition for PMMs in a system with finite Zeeman energy in the absence of Coulomb interaction. We then consider two numerical models, one in which ECT and CAR are transmitted via superconducting bulk states and one in which they are transmitted via an Andreev bound state. 
We demonstrate that the analytical sweet spot conditions can only be approximated in these more realistic models, but they cannot be satisfied exactly. As a consequence, we do not find perfect PMMs in these systems, but instead near-zero-energy states that are highly, but not perfectly, localized. These states can be considered as imperfect PMMs and their classification relies on threshold values, which adds some arbitrariness to the concept of PMMs. 

\end{abstract}

\maketitle

\begin{figure}
\centering
\includegraphics[width=0.7\linewidth]{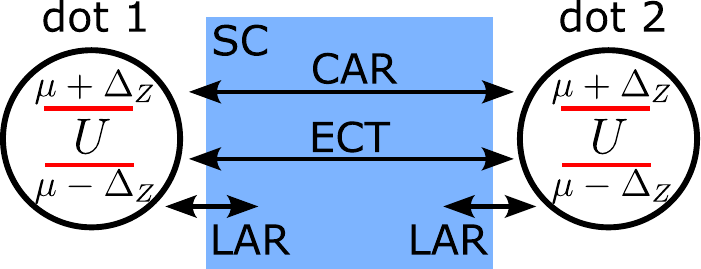}
\caption{Schematic of a minimal Kitaev chain. The minimal chain consists of two QDs with a superconductor (blue) between them. The superconductor is shorter than, or of the order of, its coherence length and therefore, crossed Andreev reflection (CAR) and elastic cotunneling (ECT) couple the QDs. Additionally, local Andreev reflection (LAR) induces superconducting pairing on the QDs. The QDs have two energy levels each, which are Zeeman split to $\mu \pm \Delta_Z$, where $\mu$ is the chemical potential and $\Delta_Z$ is the Zeeman energy. There is also an on-site Coulomb repulsion $U$ on the QDs.}
\label{fig:setup_1}
\end{figure}

\setlength{\abovedisplayskip}{6 pt}
\setlength{\belowdisplayskip}{6 pt}

\section{\label{sec:intro}Introduction}
Majorana bound states (MBSs) are quasiparticles that emerge in topological superconductors~\cite{kitaev2001unpaired} and are proposed candidates to store and manipulate quantum information in a fault-tolerant way~\cite{kitaev2003fault, nayak2008non, elliot2015colloquium} due to their non-Abelian exchange statistics~\cite{ivanov2001non}. 
They were prominently described using the so-called Kitaev chain~\cite{kitaev2001unpaired}, a spinless minimal model relying on $p$-wave superconductivity.  However, since there has not yet been any conclusive evidence for intrinsic $p$-wave superconductivity in nature, the focus has been on experimentally accessible systems that can host MBSs. One such system are nanowires with strong spin-orbit interaction (SOI) that are proximitized by a superconductor~\cite{lutchyn2010majorana, oreg2010helical, stanescu2011majorana, mourik2012signatures, das2012zero, deng2012anomalous, laubscher2021majorana}. Although these heterostructures have been studied extensively, no conclusive observation of MBSs has been made so far. One reason for this is that disorder in the nanowire can give rise to signals that mimic MBSs and this issue has been difficult to mitigate~\cite{kells2012near, lee2012zero, rainis2013towards, roy2013topologically, ptok2017controlling, liu2017andreev, moore2018two, moore2018quantized, reeg2018zero, vuik2019reproducing, stanescu2019robust, woods2019zero, chen2019ubiquitous, awoga2019supercurrent, prada2020andreev, yu2021non, sarma2021disorder, valentini2021nontopological, hess2021local,hess2022trivial, hess2022prevalence}. 

To overcome disorder and related issues, it has been proposed to implement the Kitaev chain as a chain of quantum dots (QDs)~\cite{sau2012realizing, leijnse2012parity}. It has been shown that a simple model of a minimal chain consisting of only two QDs is sufficient to obtain Majorana-like states~\cite{leijnse2012parity}. These states share most of their properties with the MBSs found in long Kitaev chains, i.e., they are at zero energy, separated from the excited states by a finite gap, and have non-Abelian exchange statistics. However, they exist only at finely tuned points of parameter space, called ``sweet spots'', and therefore they lack topological protection. As a result of this lack of protection, MBSs in minimal Kitaev chains have been dubbed ``poor man's Majoranas'' (PMMs).

In recent experiments, control over the strength of elastic cotunneling (ECT) 
and crossed Andreev reflection (CAR)~\cite{recher2001andreev}, two crucial ingredients for PMMs, has been demonstrated~\cite{wang2022singlet, bordin2023tunable, wang2023triplet, bordin2023crossed}. Consequently, there has been a large interest in studying PMMs, both theoretically~\cite{sau2012realizing, leijnse2012parity, fulga2013adaptive, tsintzis2022creating, liu2022tunable, miles2023kitaev, samuelson2023minimal, liu2023enhancing, souto2023probing, boross2023braiding, tsintzis2023roadmap, geier2023fermionparity, liu2024coupling, luna2024fluxtunable, bozkurt2024interaction} and experimentally~\cite{dvir2023realization, haaf2023engineering, zatelli2023robust, bordin2024signatures, vandriel2024crossplatform}. 
Braiding protocols for PMMs have even been proposed to use PMMs as Majorana qubits~\cite{boross2023braiding, tsintzis2023roadmap, geier2023fermionparity}.

The basic idea of PMMs in a minimal Kitaev chain is to have two QDs, connected by a superconductor; see Fig.~\ref{fig:setup_1}. Electrons on the QDs interact with each other via the superconductor through CAR and ECT~\cite{leijnse2012parity}. Assuming perfectly spin-polarized QDs, a spinless effective theory emerges, for which one can find PMMs at sweet spots where the amplitudes for CAR and ECT are equal~\cite{leijnse2012parity}. 
In the case of a finite spin-splitting on the QDs, spin has to be considered explicitly, and local Andreev reflection (LAR) is an additional interaction that affects the electrons. 
In finite spin-split minimal Kitaev chains, sweet spots in the sense of the original definition used in Ref.~\cite{leijnse2012parity}, i.e., points in parameter space allowing for perfectly localized zero-energy states, have not yet been found. 
However, parameters for which near-zero-energy states that are highly, but not perfectly, localized, have been found numerically~\cite{tsintzis2022creating, liu2023enhancing, samuelson2023minimal, souto2023probing}. 
We emphasize that in this work, we refer to a point in parameter space as a sweet spot only if the resulting states are exactly at zero energy and are perfectly localized. We refer to these states as ``perfect PMMs.'' This deviates from the usage of the term sweet spot in recent literature, see Refs.~\cite{haaf2023engineering, tsintzis2022creating, tsintzis2023roadmap, samuelson2023minimal}, where it was used to describe points in parameter space that only allow highly localized near-zero-energy states. We will henceforth label such states as ``imperfect PMMs.''

In this work, we show analytically that, even with finite spin-splitting and taking spin-degrees-of-freedom into account explicitly, a sweet spot allowing for perfect PMMs exists, if there is no on-site Coulomb interaction.
To reach this sweet spot, however, independent control over all parameters of the Hamiltonian is required. This assumption is unrealistic for more microscopic models, where the interactions between the QDs are either transmitted via superconducting bulk states~\cite{sau2012realizing, scherubl2019transport, spethmann2023highfidelity} or via an Andreev bound state (ABS)~\cite{tsintzis2022creating, liu2022tunable, miles2023kitaev, souto2023probing, liu2023enhancing, zatelli2023robust, luna2024fluxtunable}. In these more microscopic models, the sweet spot can only be approximated, resulting in imperfect PMMs within a ``threshold region'' in parameter space. These states are present even for finite Coulomb interaction. Their classification, however, depends on threshold values that measure the PMMs quality. These threshold values add some arbitrariness to the concept of PMMs.

The rest of this paper is structured as follows. We introduce a minimal spinful effective model and the quantities required to assess PMMs in Sec.~\ref{sec:setup_and_model}. In Sec.~\ref{sec:sweet_spot}, we derive a condition for sweet spots in the absence of Coulomb interaction. In Sec.~\ref{sec:reaching_sweet_spot_effective_models}, we discuss if these sweet spots can be reached in more microscopic models. We discuss if PMMs still exist in systems that have nonzero Coulomb interaction and how these are related to previously derived sweet spots in Sec.~\ref{sec:including_Coulomb}. Finally, we conclude in Sec.~\ref{sec:conclusion}. We give further information on the analytical and numerical calculations in the Appendix.

\section{\label{sec:setup_and_model}Setup and model}
The minimal system required to generate PMMs consists of two QDs with a superconducting section between them; see Fig.~\ref{fig:setup_1}. A global magnetic field $\mathbf{B}$
is applied and we choose the spin quantization axis to be parallel to the direction of the magnetic field. The coordinate system is rotated such that $\mathbf{B}$ is parallel to the $z$ axis.
Both dots have two available energy levels, at $\mu \pm \Delta_Z$, where $\mu$ is the chemical potential, measured with respect to the chemical potential of the superconductor
 and 
$\Delta_Z=g \mu_B |\mathbf{B}|/2$
(with $g$ the $g$ factor of the QDs and $\mu_B$ the Bohr magneton) the Zeeman energy. 
We assume that $\mu$ and $\Delta_Z$ are equal on both QDs, but do not expect our conclusion to depend on this assumption. We comment on the asymmetric case in Sec.~\ref{sec:sweet_spot} and in Appendix~\ref{app:asymmetric}.
The superconducting section between the two QDs is shorter than, or of the order of, the coherence length, such that CAR and ECT between the QDs are sizable~\cite{leijnse2012parity}.  
Integrating out the superconductor gives an effective theory that describes the QDs only. 
Since our model includes electron-electron interactions, it is more convenient to describe the Hamiltonian in second quantization, for which we use the basis 
\begin{align} 
	\Psi_\mathrm{even} =
	\big(
	&\ket{0,0},
	\ket{\up\dn,0}, 
	\ket{\up,\up},
	\ket{\dn,\up},
	\nonumber \\
	&\ket{\up,\dn},
	\ket{\dn,\dn},
	\ket{0,\up\dn},
	\ket{\up\dn,\up\dn}
	\big)^T, \\
	\Psi_\mathrm{odd} =
	\big(
	&
	\ket{\up,0},
	\ket{\dn,0},
	\ket{0,\up},
	\ket{\up\dn,\up}, \nonumber \\
	&\ket{0,\dn},
	\ket{\up\dn,\dn},
	\ket{\up,\up\dn},
	\ket{\dn,\up\dn}
	\big)^T,
\end{align}
where the notation is such that, e.g., $\ket{\up,\updn} = d_{1\up}^\dagger d_{2\up}^\dagger d_{2\dn}^\dagger \ket{0}$, with $d_{j\sigma}^\dagger$ creating an electron on QD $j$ with spin $\sigma$ and $\ket{0}$ is the vacuum state. The Hamiltonian conserves the particle number parity and, therefore, it is block-diagonal in the even-odd subspace. The two blocks are given by
\begin{widetext}
	\begin{subequations}  \allowdisplaybreaks
		 \label{eq:hamiltonian_second_quantization_explicit}
		\begin{align}
			&\mathcal{H}_{\mathrm{even}}  =
			\left(
			\begin{matrix}
				0 & (\Gamma^\mathrm{LAR})^* & (\Gamma^\mathrm{CAR}_{\upup,1})^* & (\Gamma^\mathrm{CAR}_{\dnup,1})^* &
				(\Gamma^\mathrm{CAR}_{\updn,1})^* & (\Gamma^\mathrm{CAR}_{\dndn,1})^* & (\Gamma^\mathrm{LAR})^* & 0 \\[4pt]
				& 2\mu + U & \Gamma^\mathrm{ECT}_{\dnup,1} & -\Gamma^\mathrm{ECT}_{\upup,1} & \Gamma^\mathrm{ECT}_{\dndn,1} &
				-\Gamma^\mathrm{ECT}_{\updn,1} & 0 & (\Gamma^\mathrm{LAR})^* \\[4pt]
				& & 2\mu + 2 \Delta_Z & 0 & 0 & 0 &
				-\Gamma^\mathrm{ECT}_{\updn,2} & -(\Gamma^\mathrm{CAR}_{\dndn,2})^* \\[4pt]
				& & & 2\mu  & 0 & 0 & -\Gamma^\mathrm{ECT}_{\dndn,2} & (\Gamma^\mathrm{CAR}_{\updn,2})^* \\[4pt]
				& & & & 2\mu  & 0 & \Gamma^\mathrm{ECT}_{\upup,2} & (\Gamma^\mathrm{CAR}_{\dnup,2})^* \\[4pt]
				& & & & & 2\mu - 2 \Delta_Z & \Gamma^\mathrm{ECT}_{\dnup,2} & -(\Gamma^\mathrm{CAR}_{\upup,2})^* \\[4pt]
				& & & & & & 2\mu + U & (\Gamma^\mathrm{LAR})^* \\[4pt]
				& & & & & & & 4\mu + 2 U \\
			\end{matrix} \right) , \\
			&\mathcal{H}_{\mathrm{odd}} = \left(
			\begin{matrix}
				\mu + \Delta_Z  & 0 & \Gamma^\mathrm{ECT}_{\upup,3} & (\Gamma^\mathrm{CAR}_{\dnup,3})^* & \Gamma^\mathrm{ECT}_{\updn,3} & (\Gamma^\mathrm{CAR}_{\dndn,3})^* & (\Gamma^\mathrm{LAR})^* & 0 \\[4pt]
				& \mu - \Delta_Z & \Gamma^\mathrm{ECT}_{\dnup,3} & -(\Gamma^\mathrm{CAR}_{\upup,3})^* & \Gamma^\mathrm{ECT}_{\dndn,3} & -(\Gamma^\mathrm{CAR}_{\updn,3})^* & 0 & (\Gamma^\mathrm{LAR})^* \\[4pt]
				& & \mu + \Delta_Z  & (\Gamma^\mathrm{LAR})^* & 0 & 0 & -(\Gamma^\mathrm{CAR}_{\updn,4})^* & -(\Gamma^\mathrm{CAR}_{\dndn,4})^* \\[4pt]
				& & & 3\mu + \Delta_Z+U & 0 & 0 & -\Gamma^\mathrm{ECT}_{\dndn,4} & \Gamma^\mathrm{ECT}_{\updn,4} \\[4pt]
				& & & & \mu - \Delta_Z & (\Gamma^\mathrm{LAR})^* & (\Gamma^\mathrm{CAR}_{\upup,4})^* & (\Gamma^\mathrm{CAR}_{\dnup,4})^* \\[4pt]
				& & & & & 3\mu - \Delta_Z + U & \Gamma^\mathrm{ECT}_{\dnup,4} & -\Gamma^\mathrm{ECT}_{\upup,4}\\[4pt]
				& & & & & & 3\mu + \Delta_Z + U & 0 \\[4pt]
				& & & & & & & 3\mu - \Delta_Z + U \\
			\end{matrix}  \right),
		\end{align}
	\end{subequations}
\end{widetext}
where $U$ is the on-site Coulomb repulsion (assumed to be equal on both dots), $\Gamma^\mathrm{LAR}$ is the LAR amplitude on the QDs, $\Gamma^\mathrm{CAR}_{\sigma\sigma',l}$ ($\Gamma^\mathrm{ECT}_{\sigma\sigma',l}$) are the CAR (ECT) amplitudes. Since these amplitudes depend on the occupation number of the QDs, there are several, labeled by the index $l$. We have only given the upper triangle of the Hamiltonian in Eq.~\eqref{eq:hamiltonian_second_quantization_explicit}. The lower triangle is determined by requiring the Hamiltonian to be hermitian.

We note a significant qualitative difference between the model given in Eq.~\eqref{eq:hamiltonian_second_quantization_explicit}, where spin is considered explicitly, and the model used, e.g., in Ref.~\cite{leijnse2012parity}, where a spinless model is considered due to an infinite Zeeman energy. In Ref.~\cite{leijnse2012parity}, the basis is given by $\Psi_\mathrm{even} = (\ket{0,0}, \ket{1,1})$ and $\Psi_\mathrm{odd} = (\ket{1,0}, \ket{0,1})$, where the notation $\ket{n_1, n_2}$ signifies a states that has $n_1 \in \{0,1\}$ ($n_2 \in \{0,1\}$) electrons on the left (right) QD. The corresponding Hamiltonian is given by
\begin{align} \label{eq:hamiltonian_spinless} 
\mathcal{H}_\mathrm{even} &= \begin{pmatrix}
0 & \Gamma^\mathrm{CAR} \\
\left(\Gamma^\mathrm{CAR}\right)^* & 2 \epsilon
\end{pmatrix}  ,
\nonumber \\ 
\mathcal{H}_\mathrm{odd} &= \begin{pmatrix}
	\epsilon & \Gamma^\mathrm{ECT} \\
	\left(\Gamma^\mathrm{ECT}\right)^* & \epsilon
\end{pmatrix} ,
\end{align}
where $\epsilon$ is the on-site energy on the QDs 
\footnote{
We note that the spinless case can be related to the spinful case if we set $\epsilon = \mu \pm \Delta_Z$ and take the limit $\Delta_Z \rightarrow \infty$.
}  
and $\Gamma^\mathrm{ECT}$ ($\Gamma^\mathrm{CAR}$) is the ECT (CAR) amplitude. The ground states in the even and odd sectors are degenerate if $| \Gamma^\mathrm{ECT}| = \sqrt{\epsilon^2+|\Gamma^\mathrm{CAR}|^2}$. In addition, if we impose the condition that there should be no charge difference on the QDs for even and odd ground states (see below), then we get that $\epsilon=0$. Here, the SOI in the bulk superconductor allows for CAR pairing. Since only CAR induces a superconducting pairing between two particles, the odd sector of the spinless model consisting of one particle does not experience any superconducting pairing. This seems to contradict the appearance of PMMs since they require superconductivity. In the spinful model given in Eq.~\eqref{eq:hamiltonian_second_quantization_explicit}, however, both the even and odd sectors contain superconducting pairing.

We label the eigenvectors of $\mathcal{H}_\mathrm{even}$ ($\mathcal{H}_\mathrm{odd}$) given in Eq.~\eqref{eq:hamiltonian_second_quantization_explicit} as $\ket{\Psi_a^\mathrm{even}}$ ($\ket{\Psi_a^\mathrm{odd}}$) and their corresponding eigenvalues as $E_a^\mathrm{even}$ ($E_a^\mathrm{odd}$), where $a$ numbers the eigenvalues, ordered such that $E_0^\mathrm{even} \leq E_1^\mathrm{even} \leq E_2^\mathrm{even} \leq \dots$ ($E_0^\mathrm{odd} \leq E_1^\mathrm{odd} \leq E_2^\mathrm{odd} \leq \dots$). 

As stated in Ref.~\cite{tsintzis2022creating}, a system that has PMMs fulfills four conditions. 
First, the even and odd parity ground states are degenerate. Second, the PMMs are chargeless. Third, the PMMs are perfectly localized with one PMM per QD. Fourth, the PMMs are separated from excited states by a finite gap. To translate these conditions to quantitative measures, we introduce the energy difference~\cite{leijnse2012parity}
\begin{equation} \label{eq:definition_dE} 
\Delta E = E_0^\mathrm{even} - E_0^\mathrm{odd},
\end{equation}
the charge difference on QD $j$~\cite{tsintzis2022creating}
\begin{equation} \label{eq:definition_dQ} 
\!\!\!\!\! \Delta Q_j \!=\! \sum_{\sigma} \! \big( \!
\bra{\Psi_0^\mathrm{even}} \! n_{j\sigma} \! \ket{\Psi_0^\mathrm{even}}
\!-\! \bra{\Psi_0^\mathrm{odd}} \! n_{j\sigma} \! \ket{\Psi_0^\mathrm{odd}}
\!
\big),
\end{equation}
where $n_{j\sigma}$ is the number of electrons on dot $j$ with spin $\sigma$,
the Majorana polarization (MP) on QD $j$~\cite{aksenov2020strong, tsintzis2022creating, samuelson2023minimal},
\begin{align} \label{eq:definition_M} 
&M_j = \frac{
\left|\sum_{\sigma} \sum_{s=\pm 1} 
\bra{\Psi_0^\mathrm{even}} \eta_{j \sigma s} \ket{\Psi_0^\mathrm{odd}}^2
\right|
}{
\sum_{\sigma} \sum_{s=\pm 1}  \left|
\bra{\Psi_0^\mathrm{even}} \eta_{j \sigma s} \ket{\Psi_0^\mathrm{odd}}^2
\right|
}, \nonumber \\
&\eta_{j\sigma +} = d_{j\sigma} + d_{j\sigma}^\dagger, \, \,
\eta_{j\sigma -} = i ( d_{j\sigma} - d_{j\sigma}^\dagger ),
\end{align}
and the excitation gap
\begin{equation} \label{eq:definition_gap} 
E_\mathrm{ex} = \min \{
E_1^\mathrm{even} - E_0^\mathrm{even},
E_1^\mathrm{odd} - E_0^\mathrm{odd}
\}.
\end{equation}
A system with perfect PMMs has $\Delta E = 0$, $\Delta Q_1=\Delta Q_2=0$, $M_1=M_2=1$, and $E_\mathrm{ex} > 0$ and the corresponding point in parameter space is called sweet spot. States with $|\Delta E| >0$, $|\Delta Q_1|, |\Delta Q_2| > 0$, $M_1, M_2 < 1$, and $E_\mathrm{ex}>0$ can still have PMM-like behavior and we call such states imperfect PMMs, which we will consider in more detail in Sec.~\ref{sec:reaching_sweet_spot_effective_models}. In contrast to recent literature, see Refs.~\cite{haaf2023engineering, tsintzis2022creating, tsintzis2023roadmap, samuelson2023minimal}, we do not call the corresponding point in parameter space sweet spot and reserve this term only for perfect PMMs, in accordance to the definition used in Ref.~\cite{leijnse2012parity}.

\section{\label{sec:sweet_spot} Analytical sweet spot condition without on-site Coulomb repulsion}

We first assume that there is no on-site Coulomb repulsion, i.e., $U=0$. In this case, the system is noninteracting, and one can equivalently describe the model given in Eq.~\eqref{eq:hamiltonian_second_quantization_explicit} using the Nambu basis
\begin{equation}
	d = (
		d_{1\up},
		d_{1\dn},
		d_{1\up}^\dagger,
		d_{1\dn}^\dagger,
		d_{2\up},
		d_{2\dn},
		d_{2\up}^\dagger,
		d_{2\dn}^\dagger
	)^T,
\end{equation}
where $d_{j\sigma}^\dagger$ ($d_{j\sigma}$) creates (annihilates) a particle of spin $\sigma$ on QD $j$.
In this basis, the Hamiltonian defined in Eq.~\eqref{eq:hamiltonian_second_quantization_explicit} becomes
\begin{subequations} \label{eq:ham_tight_binding}
\begin{align} 
& H = d^\dagger \mathcal{H}_\mathrm{Nambu} d, \\
&\mathcal{H}_\mathrm{Nambu} =
\begin{pmatrix}
	\mathcal{H}_\mu & \mathcal{H}_\mathrm{LAR} & \mathcal{H}_{T,\mathrm{ECT}} & \mathcal{H}_{T,\mathrm{CAR}}  \\
	\mathcal{H}_\mathrm{LAR}^\dagger & - \mathcal{H}_\mu & -\mathcal{H}_{T,\mathrm{CAR}} ^* & - \mathcal{H}_{T,\mathrm{ECT}}^* \\
	 \mathcal{H}_{T,\mathrm{ECT}}^\dagger & -  \mathcal{H}_{T,\mathrm{CAR}} ^T & \mathcal{H}_\mu & \mathcal{H}_\mathrm{LAR} \\
	 \mathcal{H}_{T,\mathrm{CAR}} ^\dagger & -\mathcal{H}_{T,\mathrm{ECT}}^T & \mathcal{H}_\mathrm{LAR}^\dagger & -\mathcal{H}_\mu
\end{pmatrix},
\\
&\mathcal{H}_\mu = \mu \sigma_0 + \Delta_Z \sigma_z, \\
&\mathcal{H}_\mathrm{LAR} = i \Gamma^\mathrm{LAR} \sigma_y, \\
&\mathcal{H}_{T,\mathrm{ECT}} = \begin{pmatrix}
	\Gamma^\mathrm{ECT}_\upup& \Gamma^\mathrm{ECT}_\updn \\[4 pt]
	\Gamma^\mathrm{ECT}_\dnup & \Gamma^\mathrm{ECT}_\dndn
\end{pmatrix},\\
&\mathcal{H}_{T,\mathrm{CAR}}  = \begin{pmatrix}
	\Gamma^\mathrm{CAR}_\upup & \Gamma^\mathrm{CAR}_\updn \\[4 pt]
	\Gamma^\mathrm{CAR}_\dnup & \Gamma^\mathrm{CAR}_\dndn
\end{pmatrix}.
\end{align}
\end{subequations}
We note that in the case of zero Coulomb interaction, the CAR and ECT amplitudes do not depend on the occupation number of the dots, and therefore we set $\Gamma^\mathrm{CAR/ECT}_{\sigma\sigma',l} \equiv \Gamma^\mathrm{CAR/ECT}_{\sigma\sigma'}$ for $l=1,2,3,4$.

In this description, perfect PMMs correspond to eigenvectors with zero energy and of the form:
\begin{align} \label{eq:v1_v2_ansatz_pmm}
	v_1 &= (a_{1\up}, a_{1\dn}, a_{1\up}^*, a_{1\dn}^*, 0, 0, 0, 0)^T ,
	\nonumber \\
	v_2 &= (0, 0, 0, 0, a_{2\up}, a_{2\dn}, a_{1\up}^*, a_{2\dn}^* )^T,
\end{align} 
with $a_{j\sigma} \in \mathbb{C}$. To find conditions for the sweet spots, we thus solve the coupled system of equations $\mathcal{H}_\mathrm{Nambu} v_1 = 0$ and $\mathcal{H}_\mathrm{Nambu} v_2 = 0$.
We emphasize that if Eq.~\eqref{eq:ham_tight_binding} has two zero-energy states, then the even and odd ground states of Eq.~\eqref{eq:hamiltonian_second_quantization_explicit} are equal in energy, i.e., $\Delta E= 0$. If these zero-energy states have the form given in Eq.~\eqref{eq:v1_v2_ansatz_pmm}, then $M_1=M_2=1$. This relation between the solutions of Eqs.~\eqref{eq:hamiltonian_second_quantization_explicit} and~\eqref{eq:ham_tight_binding}, however, can only be drawn if $U=0$.

Due to gauge invariance of the superconducting term, we assume that $\Gamma^\mathrm{LAR} \in \mathbb{R}$. However, the CAR and ECT amplitudes can still have a complex phase that originates from SOI, the effect of which we can describe by the unitary operator~\cite{spethmann2023highfidelity}
\begin{equation} \label{eq:soi_matrix}
	U_\text{SOI}( \Phi_\mathrm{SOI}) = 
	\cos (\Phi_\mathrm{SOI}) 
	+ i \sin ( \Phi_\mathrm{SOI}) 
	\mathbf{n} \cdot  \boldsymbol{\sigma},
\end{equation}
where $\Phi_\mathrm{SOI}$ is the SOI angle (depending on both the strength of SOI and the distance between the QDs), for which we assume, without loss of generality, $0\leq \Phi_\mathrm{SOI} \leq \pi/2$, $\mathbf{n}=(n_x,n_y,n_z)$ is a unit vector indicating the direction of SOI, and $\boldsymbol{\sigma} = (\sigma_x, \sigma_y, \sigma_z)$ is the vector of Pauli matrices. In our system, we can always rotate the coordinate system such that $\mathbf{n}$ lies within the $yz$ plane 
and we can, therefore, set $\mathbf{n}=(0, \sin\theta,\cos\theta)$, with $\theta \in [0,\pi/2]$ the angle between the SOI direction and the magnetic field. Therefore, we can also write
\begin{align} 
U_\mathrm{SOI} (\Phi_\mathrm{SOI}) = 
\begin{pmatrix}
C e^{i\rho} & K \\
- K & C e^{-i\rho}
\end{pmatrix} ,
\end{align}
where we have defined
\begin{align} \label{eq:definition_C_rho_K}
	C e^{i \rho} &= \cos(\Phi_\mathrm{SOI}) + i \cos (\theta) \sin (\Phi_\mathrm{SOI}),
	\nonumber \\
K &= \sin (\theta) \sin (\Phi_\mathrm{SOI}),
\end{align}
with $C, K, \rho \in \mathbb{R}$. 
We will later calculate the complex phases of the ECT and CAR amplitudes using second-order perturbation theory in the hopping strength (see Sec.~\ref{sec:reaching_sweet_spot_effective_models} and Appendix~\ref{app:effective_theory}). The ECT and CAR amplitudes calculated in this fashion receive their complex phases only through SOI and motivated by their explicit expression, we use the following general ansatz for the ECT and CAR submatrices:
\begin{subequations}\label{eq:t_and_delta_explicit}
	\begin{align}
		&\mathcal{H}_{T,\mathrm{ECT}} = 
		\begin{pmatrix}
			\Gamma^\mathrm{ECT}_\upup e^{i\rho}& \Gamma^\mathrm{ECT}_\updn \\[4 pt]
			-\Gamma^\mathrm{ECT}_\updn & \Gamma^\mathrm{ECT}_\dndn e^{-i\rho}
		\end{pmatrix}, \\
		&\mathcal{H}_{T,\mathrm{CAR}}  =
		\begin{pmatrix}
			\Gamma^\mathrm{CAR}_\upup & \Gamma^\mathrm{CAR}_\updn e^{i\rho} \\[4 pt]
			- \Gamma^\mathrm{CAR}_\updn e^{-i\rho} & \Gamma^\mathrm{CAR}_\dndn
		\end{pmatrix},
	\end{align}
\end{subequations}
with $\Gamma^\mathrm{ECT/CAR}_{\sigma\sigma'} \in \mathbb{R}$ for all $\sigma,\sigma' \in \{\up, \dn\}$.
We note that if $\theta=0$, i.e., SOI is parallel  to the magnetic field, 
	then the following
	explicitly calculated ECT and CAR amplitudes in Appendix \ref{app:effective_theory} are zero: $\Gamma^\mathrm{CAR}_\upup = \Gamma^\mathrm{CAR}_\dndn=0$ and $\Gamma^\mathrm{ECT}_\updn=0$. To avoid this, we assume that $\theta>0$, i.e., SOI must not be parallel  to the magnetic field \footnote{We note that if SOI is parallel to the magnetic field, i.e., $\theta=0$, and we set $\Gamma^\mathrm{CAR}_\upup = \Gamma^\mathrm{CAR}_\dndn=0$ and $\Gamma^\mathrm{ECT}_\updn=0$ in the ansatz, there exist perfectly localized zero-energy solutions at the sweet spot defined by Eq.~\eqref{eq:sweet_spot_condition}, but they are four-fold degenerate.
	}.
Next, we introduce the unitary matrix
\begin{align}
	\! M \! = \! \mathrm{diag}  \big( \!
	e^{i\frac{\rho}{2}},
	e^{-i\frac{\rho}{2}},
	e^{-i\frac{\rho}{2}},
	e^{i\frac{\rho}{2}},
	e^{-i\frac{\rho}{2}},
	e^{i\frac{\rho}{2}},
	e^{i\frac{\rho}{2}},
	e^{-i\frac{\rho}{2}} 
	\big) .
\end{align}
The transformed Hamiltonian $\mathcal{H}^\prime_\mathrm{Nambu} = M^\dagger \mathcal{H}_\mathrm{Nambu} (\rho) M = \mathcal{H}_\mathrm{Nambu} (\rho=0)$ is real, which simplifies the search for PMMs, since we can now, without loss of generality, restrict ourselves to the case in which $\theta=\pi/2$, i.e., where the direction of the SOI is perpendicular to the magnetic field. 
Assuming that $\mu \neq \pm \Delta_Z$, we find the following sweet spot conditions allowing for perfect PMMs:
\begin{subequations} \label{eq:sweet_spot_condition}
\begin{align} 
&\mu^2-\Delta_Z^2 + \left(\Gamma^\mathrm{LAR}\right)^2 = 0,
\label{eq:sweet_spot_LAR_condition}
 \\
&\!\! \left(\mu \! + \! \Delta_Z \right)  \! \left(\Gamma^\mathrm{CAR}_\updn \! - \! s \Gamma^\mathrm{ECT}_\updn \right) \! = \! \Gamma^\mathrm{LAR} \! \left(\Gamma^\mathrm{ECT}_\upup \! - \! s \Gamma^\mathrm{CAR}_\upup \right) \!\! , \label{eq:sweet_spot_updn_condition}
\\
&\!\! \left(\mu \! + \! \Delta_Z \! \right) \! \left( \Gamma^\mathrm{CAR}_\dndn \!\! - \!\! s \Gamma^\mathrm{ECT}_\dndn \right)  \!\! = \!\! \left(\mu \! - \! \Delta_Z \!\right) \!\! \left( \Gamma^\mathrm{CAR}_\upup \!\! - \!\! s \Gamma^\mathrm{ECT}_\upup\right) \!\! ,
\label{eq:sweet_spot_dndn_condition}
\end{align}
\end{subequations}
where $s= 1$ or $s=-1$ gives us two possible sweet spot conditions.
The special case $\mu = \pm \Delta_Z$ gives the conditions $\Gamma^\mathrm{LAR}=0$, $\Gamma^\mathrm{CAR}_\updn = s\Gamma^\mathrm{ECT}_\updn$, and $\Gamma^\mathrm{CAR}_\dndn = s \Gamma^\mathrm{ECT}_\dndn$
($\Gamma^\mathrm{CAR}_\upup = s \Gamma^\mathrm{ECT}_\upup$)
 if $\mu=\Delta_Z$ ($\mu=-\Delta_Z$ ). Thus, if $\mu = \pm \Delta_Z$, then these conditions resemble the sweet spot condition found for the spinless case~\cite{leijnse2012parity}.
 In principle, a sweet spot is also possible if the chemical potentials, Zeeman energies, and LAR amplitudes on the two QDs are different. If $\mu_j$, $\Delta_{Z,j}$, and $\Gamma^\mathrm{LAR}_j$ describe the parameters on QD $j=1,2$, then a sweet spot exists if Eq.~\eqref{eq:sweet_spot_LAR_condition} is fulfilled for both QDs individually, and $\Gamma^\mathrm{CAR}_{\sigma \sigma^\prime} = s \Gamma^\mathrm{ECT}_{\sigma \sigma^\prime}$  for $\sigma \sigma^\prime \in \{ \upup, \dndn, \updn\}$. In the following we will, however, assume that $\mu_1=\mu_2 = \mu$ and $\Delta_{Z,1}=\Delta_{Z,2}=\Delta_Z$, but comment further on the asymmetric case in Appendix~\ref{app:asymmetric}. 

\begin{figure}
	\centering
	\includegraphics[width=\linewidth]{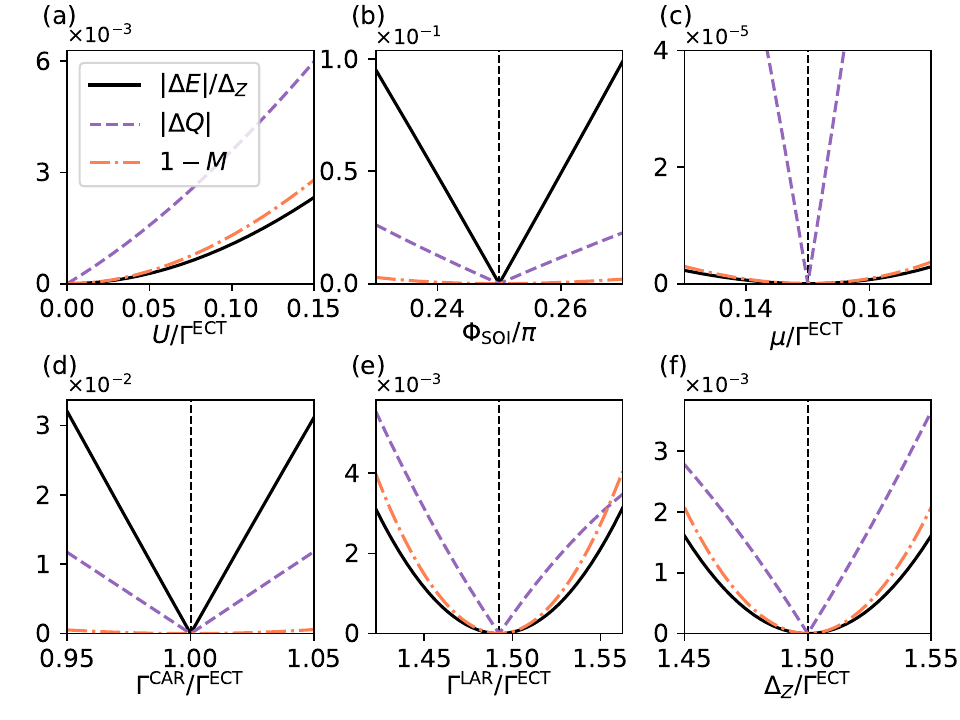}
	\caption{
	Stability of the analytic sweet spot against various parameter changes.
To calculate this data, we use Eq.~\eqref{eq:hamiltonian_second_quantization_explicit}, but assume that none of the parameters depend on the occupation of the QDs and take strong assumptions on the dependence between the parameters.
		We set $\theta = \pi/2$ and assume $\Gamma^\mathrm{ECT}_\upup=\Gamma^\mathrm{ECT}_\dndn=\Gamma^\mathrm{ECT} \cos (\Phi_\mathrm{SOI})$, $\Gamma^\mathrm{ECT}_\updn = \Gamma^\mathrm{ECT} \sin (\Phi_\mathrm{SOI})$, 
		$\Gamma^\mathrm{CAR}_\upup=\Gamma^\mathrm{CAR}_\dndn=\Gamma^\mathrm{CAR} \sin (\Phi_\mathrm{SOI})$, $\Gamma^\mathrm{CAR}_\updn = \Gamma^\mathrm{CAR} \cos (\Phi_\mathrm{SOI})$, which changes the conditions given in Eqs.~\eqref{eq:sweet_spot_updn_condition} and~\eqref{eq:sweet_spot_dndn_condition} to $\Gamma^\mathrm{ECT} = s \Gamma^\mathrm{CAR}$ and $\Phi_\mathrm{SOI} = \pi/4$. 
		The parameters used for the sweet spot are $\Gamma^\mathrm{CAR}/\Gamma^\mathrm{ECT}=1$, $\mu/\Gamma^\mathrm{ECT}=0.15$, $\Delta_Z/\Gamma^\mathrm{ECT}=1.5$, $\Gamma^\mathrm{LAR}=\sqrt{\Delta_Z^2-\mu^2}$, $U=0$, and $\Phi_\mathrm{SOI}=\pi/4$.
		In each panel, one of these parameters is tuned away from the sweet spot value.
		(a): On-site Coulomb repulsion $U$. 
		(b): SOI angle $\Phi_\mathrm{SOI}$.
		(c): Chemical potential $\mu$.
		(d): CAR amplitude $\Gamma^\mathrm{CAR}$.
		(e): LAR amplitude $\Gamma^\mathrm{LAR}$.
		(f): Zeeman energy $\Delta_Z$.
		In all cases, $|\Delta E|$, $|\Delta Q|$, and $1-M$ increase. In some cases, this increase is linear, whereas in other cases the increase is of higher power, i.e., at most quadratic. 
		We find that $|\Delta Q|$ increases linearly in all panels, $1-M$ increases at most quadratically in all panels, and $|\Delta E|$ increases linearly in panels~(b) and~(d), and at most quadratically in all other panels.
	}
	\label{fig:deviation_various_parameters_1}
\end{figure}

The field operators for the PMMs corresponding to the sweet spot found in Eq.~\eqref{eq:sweet_spot_condition} in the basis of the transformed Hamiltonian $\mathcal{H}^\prime_\mathrm{Nambu}$ are
\begin{subequations}\label{eq:wf_PMM_SOI_general} 
\begin{align}
\!\!\! \gamma_1 \!=& 
\frac{1}{\sqrt{N}} \!\! 
\left[ \!
\frac{\Gamma^\mathrm{LAR}}{\mu \!+\! \Delta_Z } \! (
e^{i\alpha} c_{1\up} \!+\! e^{-i\alpha} c_{1\up}^\dagger
) 
\!-\! ( e^{-i\alpha} c_{1\dn} \!+\! e^{i\alpha} c_{1\dn}^\dagger ) \!
\right] \!\! , \\
\!\!\! \gamma_2 \!=& 
\frac{i}{\sqrt{N}} \!\! 
\left[ \!
\frac{\Gamma^\mathrm{LAR}}{\mu \!+\! \Delta_Z } \! (
e^{i\alpha} c_{2\up} \!-\! e^{-i\alpha} c_{2\up}^\dagger
) 
\!+\! (e^{-i\alpha} c_{2\dn} \!-\! e^{i\alpha} c_{2\dn}^\dagger ) \!
\right] \!\! ,
\end{align}
\end{subequations}
where $N=2[1+(\Gamma^\mathrm{LAR})^2/(\mu+\Delta_Z)^2]$ and $\alpha=0$ ($\alpha=\pi/2$) if $s=1$ ($s=-1$).
In both cases, $\gamma_j = \gamma_j^\dagger$ for $j = 1,2$ (recall that ${\Gamma^\mathrm{LAR}}$ is real) and both $\gamma_j$ are localized on one QD only. This means that one obtains $\Delta E = 0$, 
$\Delta Q_1 = \Delta Q_2 = 0$, 
$M_1 = M_2 = 1$, 
and $E_\mathrm{ex}>0$, i.e., perfect PMMs.
We comment on the size of the excitation gap at the sweet spot in Appendix~\ref{app:gap}.

As one tunes away from the sweet spot, $|\Delta E|$, $|\Delta Q_j|$, and $1-M_j$ increase; see Fig.~\ref{fig:deviation_various_parameters_1}. 
This increase is in some cases linear, in other cases slower, i.e., at most quadratic. 
This hints at some stability of the sweet spot. 
The spinless model, given in Eq.~\eqref{eq:hamiltonian_spinless} has three parameters, and $|\Delta E|$ increases quadratically if $\epsilon$ varies, but linearly if the CAR or ECT amplitudes are varied. In contrast, the spinful model given in Eq.~\eqref{eq:hamiltonian_second_quantization_explicit} has 36 free parameters, nine if $U=0$ and if one uses the ansatz given in Eq.~\eqref{eq:t_and_delta_explicit} with $\rho=0$, or six if one uses the even simpler model used for Fig.~\ref{fig:deviation_various_parameters_1}. In any case, the spinful model naturally has more parameters than the spinless model, which increases the parameter space significantly, thus making the search for a sweet spot more complicated, requiring more fine-tuning of parameters, and possibly leading to a more unstable sweet spot. We have, however, shown that a sweet spot exists for the spinful model, and that it is not highly unstable, since $|\Delta E|$ increases linearly only for some parameters, while increasing at most quadratically for other parameters. However, the increase of $|\Delta Q_j|$ was found to be linear for all parameter variations.

\section{\label{sec:reaching_sweet_spot_effective_models} PMMs in microscopic theories}
We have analytically derived a condition for sweet spots in Eq.~\eqref{eq:sweet_spot_condition}.
If the ECT, CAR, and LAR amplitudes were independently tunable, these conditions could be fulfilled and one would find perfect PMMs. 
In reality, however, ECT, CAR, and LAR are effective processes and their amplitudes depend on other parameters. The amplitudes also depend on how these processes are transmitted, as this can happen either through superconducting bulk states or through ABSs in the superconducting region. In this section, we use more realistic models to study these two cases  and demonstrate that by using these models, the sweet spot conditions can only be approximately satisfied and the resulting states are imperfect PMMs. We emphasize that for now we still assume zero on-site Coulomb repulsion $U$. We will consider nonzero Coulomb repulsion in Sec.~\ref{sec:including_Coulomb}.

\subsection{Transmission via superconducting bulk states}

\begin{figure}
	\centering
	\includegraphics[width=0.7\linewidth]{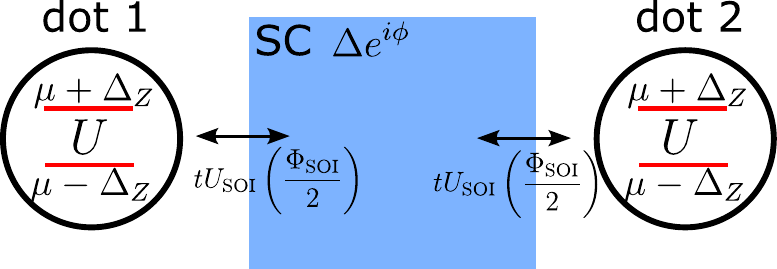}
	\caption{Model of CAR and ECT transmitted by superconducting bulk states. The superconductor that transmits effective couplings between the two outer QDs is considered explicitly. The superconductor has a gap $\Delta$ and a phase $\phi$. The hopping between the QDs and the superconductor is characterized by the amplitude $t$ and the SOI matrix $U_\mathrm{SOI}(\Phi_\mathrm{SOI}/2)$.}
	\label{fig:setup_3}
\end{figure}

If the effective coupling between the two QDs is transmitted via superconducting bulk states~\cite{sau2012realizing, leijnse2012parity}, the setup is given by Fig.~\ref{fig:setup_3} and the full Hamiltonian has the form~\cite{scherubl2019transport, spethmann2023highfidelity}
\begin{align} 
H_\mathrm{bulk} =& \sum_{j=1,2} \Big( \sum_{\sigma=\up,\dn} \frac{\mu + \sigma \Delta_Z}{2} d_{j\sigma}^\dagger d_{j\sigma} + \frac{U}{2} d_{j\up}^\dagger d_{j\up} d_{j\dn}^\dagger d_{j\dn} \Big)
\nonumber \\ &+ 
\sum_{\mathbf{k}} \Big[ \sum_{\sigma=\up,\dn} \frac{\epsilon_\mathbf{k}}{2} c_{\mathbf{k}\sigma}^\dagger c_{\mathbf{k} \sigma} 
- \Delta e^{-i\phi} c_{\mathbf{k}\up}^\dagger c_{-\mathbf{k}\dn}^\dagger
 \nonumber \\ 
& +
t \!\! \sum_{\sigma,\sigma'=\up,\dn} \!\! U_\mathrm{SOI} \left(\frac{\Phi_\mathrm{SOI}}{2}\right)_{\sigma\sigma'} \!\!
(c_{\mathbf{k}\sigma}^\dagger d_{1\sigma'} 
+ d_{2\sigma}^\dagger c_{\mathbf{k}\sigma})
\Big] 
\nonumber \\ & +
\text{H.c.},
\label{eq:hamiltonian_sc_bulk_full}
\end{align}
where $d_{j\sigma}^\dagger$ ($d_{j\sigma}$) creates (annihilates) an electron with spin $\sigma$ on QD $j$, $c_{\mathbf{k}\sigma}^\dagger$ ($c_{\mathbf{k}\sigma}$) creates (annihilates) an electron with momentum $\mathbf{k}$ and spin $\sigma$ in the superconductor, $\epsilon_\mathbf{k}$ is the energy of an electron with momentum $\mathbf{k}$,
 $\Delta \in \mathbb{R}$ is the superconducting gap, $\phi$ is the complex phase of the superconducting pairing, $U_\mathrm{SOI} (\Phi_\mathrm{SOI}/2)$ is the SOI matrix defined in Eq.~\eqref{eq:soi_matrix}, and the notation $\mu + \sigma \Delta_Z$ means $\mu + \Delta_Z$ ($\mu-\Delta_Z$) if $\sigma=\up$ ($\sigma=\dn$). In this section, we set the direction of SOI to be $\mathbf{n} = (0,1,0)$ and the superconducting phase to be $\phi=0$.
We use Schrieffer-Wolff perturbation theory to integrate out the superconductor~\cite{choi2000spin, winkler2003spin, bravyi2011schrieffer, probst2016signatures, schrade2017detecting, scherubl2019transport, spethmann2022coupled, spethmann2023highfidelity}; see Appendix~\ref{app:effective_theory}.
The chemical potential $\mu$ and the Zeeman energy $\Delta_Z$ are renormalized by coupling the QDs to the superconductors. Furthermore, we obtain analytical expressions for the ECT, CAR, and LAR amplitudes.
We emphasize that the effective parameters are integral expressions that only converge if $\mu$, $\Delta_Z$, and $U$ lie within a certain region of parameter space (see Appendix~\ref{app:effective_theory}). In all calculations presented here, the parameters are within this region of parameter space and therefore the corresponding integrals converge.

\begin{figure}
	\centering
	\includegraphics[width=\linewidth]{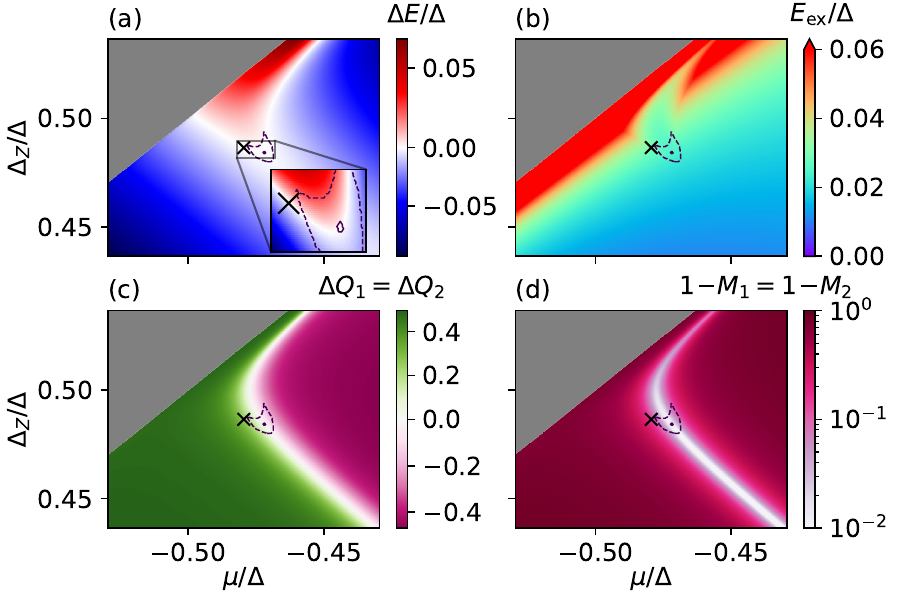}
	\caption{Threshold regions (TRs) for a model in which CAR and ECT are transmitted via superconducting bulk states, shown in Fig.~\ref{fig:setup_3}. We vary the chemical potential $\mu$ and the Zeeman energy $\Delta_Z$ to search for a TR. In panel~(a), the energy difference $\Delta E$ between the even and odd ground states is shown, the excitation gap $E_\mathrm{ex}$ is shown in panel~(b), panel~(c) shows the charge difference $\Delta Q_j$, and the MP $M_j$ is shown in panel~(d). Because the system is symmetric, we obtain $\Delta Q_1=\Delta Q_2$ and $M_1=M_2$. 
		The gray region indicates the area of parameter space for which the integrals for the effective parameters diverge; see Appendix~\ref{app:effective_theory}.
		The black cross indicates the numerically found approximate solution for the sweet spot condition given in Eq.~\eqref{eq:sweet_spot_condition} (see Appendix~\ref{app:sweet_spot_analytics} for more detail). This optimized solution is close to the TR, which is indicated by the black circled area. We show two TRs with different threshold values: a dashed line with loose threshold values and a solid line for a TR with stricter threshold values [better visible in zoomed-in inset in panel~(a)]. 
		This figure was calculated using the effective Hamiltonian given in Appendix~\ref{app:effective_theory} with the parameters 
		$U=0$, $t/\Delta = 0.1$, $\Phi_\mathrm{SOI}=0.1\pi$. The threshold values for the dashed TR are 	
		$\Delta E_\mathrm{th}/\Delta = 10^{-3}$, $\Delta Q_\mathrm{th}=0.2$, and $M_\mathrm{th}=0.2$ 
		and the threshold values for the solid TR are ~
		$\Delta E_\mathrm{th}/\Delta = 10^{-4}$, $\Delta Q_\mathrm{th}=0.02$, and $M_\mathrm{th}=0.02$. Note, throughout we do not optimize for the excitation gap, $E_\mathrm{ex}$, but check afterwards that it is finite within the TR. 
		We emphasize that the color scale in the inset of panel~(a) is different than in the main part of (a). The values in the inset vary between $\Delta E / \Delta = 5 \times 10^{-3}$ (blue), $\Delta E = 0$ (white), and $\Delta E / \Delta = 3 \times 10^{-3}$ (red).
	}
	\label{fig:analytical_sweet_spot_bulk}
\end{figure}

We vary $\mu$ and $\Delta_Z$ to find a solution for the sweet spot condition given in Eq.~\eqref{eq:sweet_spot_condition}. However, we do not find an exact solution to the sweet spot condition and up to numerical accuracy, we do not find perfect PMMs with $\Delta E=0$, $\Delta Q_1 = \Delta Q_2 = 0$, $M_1 = M_2 = 1$, and $E_\mathrm{ex} > 0$.
Nevertheless, the best approximate solution for Eq.~\eqref{eq:sweet_spot_condition} (see Appendix~\ref{app:sweet_spot_analytics}) is close to a region in parameter space that has imperfect PMMs, i.e., highly, but not perfectly, localized near-zero-energy states; see Fig.~\ref{fig:analytical_sweet_spot_bulk}.
To characterize the imperfect PMMs, we introduce the notion of a threshold region (TR) as a region of parameter space in which
\begin{align} \label{eq:definition_ROT}
	&|\Delta E | < \Delta E_\mathrm{th}
	\text{ and }
	E_\mathrm{ex} > E_\mathrm{ex, th} 
	\nonumber \\ &
	\text{ and }
	|\Delta Q_1 | < \Delta Q_\mathrm{th}
	\text{ and }
	|\Delta Q_2 | < \Delta Q_\mathrm{th}
	\nonumber \\ &
	\text{ and }
	M_1 > 1-M_\mathrm{th}
	\text{ and }
	M_2 > 1-M_\mathrm{th},
\end{align}
where $\Delta E_\mathrm{th}$, $E_\mathrm{ex, th} $, $\Delta Q_\mathrm{th}$, and $M_\mathrm{th}$ are threshold values to be chosen. 
States that do not satisfy the TR condition given in Eq.~\eqref{eq:definition_ROT} are not classified as PMMs.
We stress that the concept of imperfect PMMs and threshold values to classify PMMs has been introduced before; see, e.g., Refs.~\cite{tsintzis2022creating, haaf2023engineering, souto2023probing, tsintzis2023roadmap, samuelson2023minimal}. We emphasize, however, the important distinction between perfect and imperfect PMMs, which has not been done in recent literature~\cite{haaf2023engineering, tsintzis2022creating, tsintzis2023roadmap, samuelson2023minimal}, and note that due to allowing imperfect PMMs, the classification of PMMs is, to a certain extent, arbitrary since it depends on threshold values. Some limits on the threshold values can be set, e.g., by braiding. In Ref.~\cite{tsintzis2023roadmap}, braiding was studied for $M_\mathrm{th} \approx 10^{-2} \dots 10^{-4}$ and in Ref.~\cite{haaf2023engineering} the MP of PMMs was experimentally estimated to be above $0.9$. The energy difference $\Delta E$ was limited to be $\Delta E/\Delta \approx 10^{-3} \dots 10^{-2}$ in Refs.~\cite{tsintzis2023roadmap, samuelson2023minimal}. The threshold values chosen in this work are consistent with these values.

\subsection{Transmission via ABSs}

\begin{figure}
	\centering
	\includegraphics[width=0.7\linewidth]{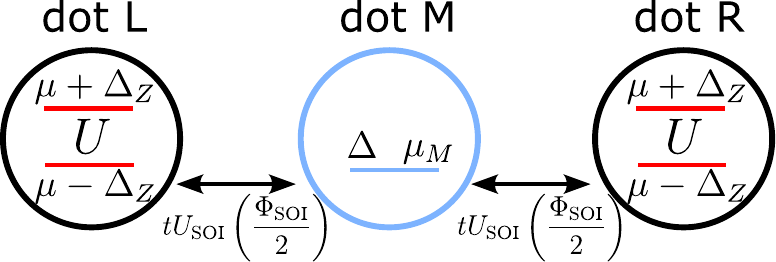}
	\caption{Transmission of CAR and ECT via an Andreev bound state (ABS). The ABS that transmits the effective couplings is considered explicitly, in the form of an additional QD (blue), placed between the two outer QDs. The left and right QDs are equivalent to the two QDs in the setup shown in Fig.~\ref{fig:setup_1}. The middle QD has a chemical potential $\mu_M$ and a superconducting pairing potential $\Delta$. Due to screening effects of the superconductor, we assume that the Zeeman energy and Coulomb repulsion on the middle QD are zero. The hopping between the QDs is characterized by the amplitude $t$ and the SOI angle $\Phi_\mathrm{SOI}$.}
	\label{fig:setup_ABS}
\end{figure}

If the superconducting section between the two QDs is composed of a semiconducting section that is proximitized by a superconductor, then this hybrid section hosts ABSs which can then transmit an effective coupling between the QDs~\cite{tsintzis2022creating, dvir2023realization}. A simplified model of this setup was considered in Ref.~\cite{tsintzis2022creating}, where an ABS was modeled as an additional QD between the two outer QDs; see Fig.~\ref{fig:setup_ABS}. The Hamiltonian for this setup is given by~\cite{tsintzis2022creating}
\begin{align} \label{eq:hamiltonian_ABS_simple}
	H_\mathrm{ABS} =& \sum_{j=1,2} \Big( 
	\sum_{\sigma=\up,\dn}
	\frac{\mu + \sigma \Delta_Z}{2} d_{j\sigma}^\dagger d_{j\sigma}  
	+  \frac{U}{2} d_{j\up}^\dagger d_{j\up}  d_{j\dn}^\dagger d_{j\dn}  
	\Big)
	\nonumber \\ &+ 
	\sum_{\sigma=\up,\dn} \frac{\mu_M}{2} c_\sigma^\dagger c_\sigma 
	- \Delta c_\up^\dagger c_\dn^\dagger
	\nonumber \\ &+ 
	t \sum_{\sigma,\sigma'=\up,\dn}  
	U_\mathrm{SOI} \left(\frac{\Phi_\mathrm{SOI}}{2}\right)_{\sigma \sigma'}
	(c_\sigma^\dagger d_{1\sigma'} + d_{2\sigma}^\dagger c_{\sigma'} )  
	\nonumber \\ &+ 
	\mathrm{H.c.},
\end{align}
where the only difference to Eq.~\eqref{eq:hamiltonian_sc_bulk_full} is that $c_{\sigma}^\dagger$ ($c_{\sigma}$) creates (annihilates) a particle of spin $\sigma$ on the central QD, $\mu_M$ is the chemical potential on the center QD, and there is no summation over momenta. Furthermore, we assume that there is no Zeeman energy and no Coulomb repulsion on the central QD due to screening by the superconductor~\cite{tsintzis2022creating}. Compared to the model introduced in Sec.~\ref{sec:via_bulk_states}, the chemical potential $\mu_M$ of the central dot adds an additional tuning knob, and like in the previous model, the chemical potential $\mu$ of the normal conducting QD also acts as a tuning knob. 
We again set $\mathbf{n}=(0,1,0)$. 
We note that rotating the magnetic field, which is equivalent to tuning $\mathbf{n}$, could be used to tune the system to a TR~\cite{liu2022tunable, bordin2023tunable, wang2023triplet, miles2023kitaev, haaf2023engineering, liu2024coupling}, or it might even be helpful in maximizing the excitation gap, which is, however, not studied in this work. 

\begin{figure}
	\centering
	\includegraphics[width=\linewidth]{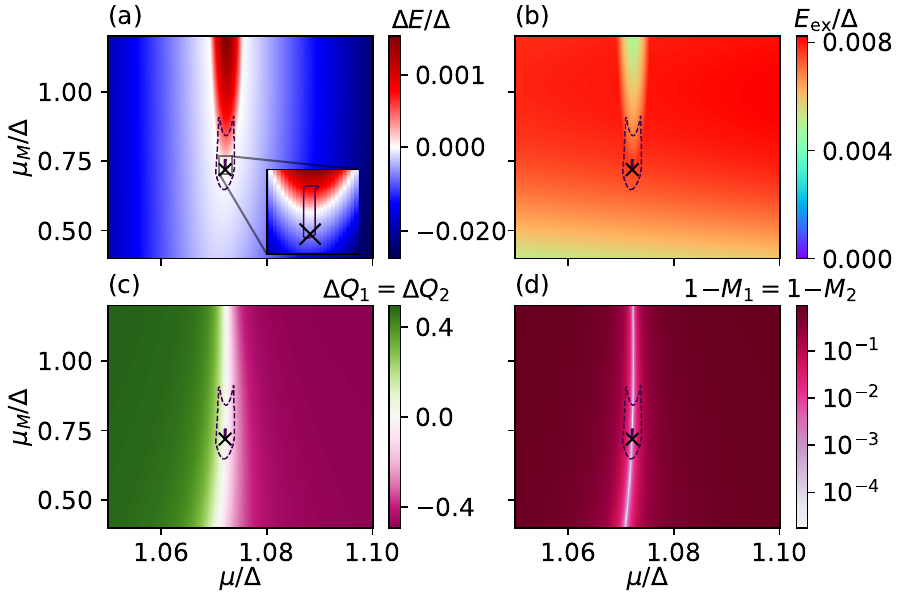}
	\caption{Threshold regions (TR) for a model in which CAR and ECT are transmitted via an ABS, shown in Fig.~\ref{fig:setup_ABS}. We vary the chemical potential $\mu$ of the two outer QDs and the chemical potential $\mu_M$ of the central QD to search for TR. In panel~(a), the energy difference $\Delta E$ between the even and odd ground states is shown, the excitation gap $E_\mathrm{ex}$ is shown in panel~(b), panel~(c) shows the charge difference $\Delta Q_j$ and the MP $M_j$ is shown in panel~(d). Because the system is symmetric, we obtain $\Delta Q_1 = \Delta Q_2$ and $M_1=M_2$. 
	The black cross indicates the parameters that solve the sweet spot condition given in Eq.~\eqref{eq:sweet_spot_condition} to second order in $t/\Delta$ [see Appendix~\ref{app:sweet_spot_analytics} for more detail], which coincides with the TR in the full model. We show two TR with different threshold values: a dashed line with loose threshold values and a solid line for a TR with stricter threshold values [better visible in the inset in panel~(a)].
		For this data, the Hamiltonian given in Eq.~\eqref{eq:hamiltonian_ABS_simple} is used with the parameters (rounded to four significant digits) $U = 0$, $\Delta_Z/\Delta = 1.0674$, 
		$t/\Delta = 0.1$, and $\Phi_\mathrm{SOI} = 0.2 \pi$, 
		$s=1$. 
		The threshold values for the dashed TR are $\Delta E_\mathrm{th}/\Delta = 5 \times 10^{-4}$,  $\Delta Q_\mathrm{th} = 0.2$, and $M_\mathrm{th} = 0.2$ and the threshold values for the solid TR are 
		$\Delta E_\mathrm{th}/\Delta = 10^{-4}$,  $\Delta Q_\mathrm{th} = 0.02$, and $M_\mathrm{th} = 0.02$. The excitation gap $E_\mathrm{ex}$ is finite in the TR. We emphasize that the color scale in the inset of panel~(a) is different than in the main part of panel~(a). The values in the inset vary between $\Delta E/\Delta = -3.9\times 10^{-4}$ (blue), $\Delta E  = 0$ (white), and $\Delta E /\Delta = 1.6 \times 10^{-4}$ (red). Furthermore, we checked that also for larger Zeeman fields ($\Delta_Z/\Delta = 2.0$), the TR is close to the analytical sweet spot and the numerically calculated states are not perfect but imperfect PMMs.
	}
	\label{fig:sweet_spot_analytics_ABS}
\end{figure}

As already done in Sec.~\ref{sec:via_bulk_states}, a Schrieffer-Wolff transformation is used on Eq.~\eqref{eq:hamiltonian_ABS_simple} to obtain an effective two-site Hamiltonian to second order in the hopping amplitude $t$, which we do in Appendix~\ref{app:effective_theory}. Since the corresponding effective parameters do not contain any integral expressions, solving the sweet spot condition defined in Eq.~\eqref{eq:sweet_spot_condition} is easier and there are no convergence-related limits on the parameters. 
The effective parameters are calculated only up to second order in the hopping amplitude $t$ and therefore, we consider the sweet spot conditions only up to second order in $t$, too. Fixing $t/\Delta$ and $\Phi_\mathrm{SOI}$, we solve the system of three coupled equations for $\mu/\Delta$, $\mu_M/\Delta$, and $\Delta_Z/\Delta$; see Appendix~\ref{app:sweet_spot_analytics}.
We use these parameters for the full numerical model given in Eq.~\eqref{eq:hamiltonian_ABS_simple} and within numerical accuracy, do not find any perfect PMMs. However, the second-order solution to Eq.~\eqref{eq:sweet_spot_condition} coincides with a TR in the full model; see Fig.~\ref{fig:sweet_spot_analytics_ABS}. 
We emphasize that although $1-M$ becomes quite small in the full model, it is still finite, of the order of $10^{-5}$, and therefore this is an imperfect PMMs.

We emphasize that, in this section, $t/\Delta$ must be small because we solve the sweet spot condition only up to second order in $t/\Delta$. As a consequence, the corresponding excitation gap $E_\mathrm{ex}$ is small too, but finite. 
The relation between $E_\mathrm{ex}$ and the hopping amplitude has been studied in Ref.~\cite{liu2023enhancing}, which demonstrated that $E_\mathrm{ex}$ grows with $t$. 
We will introduce larger values for $t/\Delta$ in Sec.~\ref{sec:including_Coulomb}.
We furthermore note that we consider only the sweet spot appearing for $\Delta_Z/\Delta>0$, $\mu/\Delta >0$, and $\mu_M/\Delta > 0$. However, there is a TR in each quadrant of the $\mu$ versus $\mu_M$ parameter space; see, e.g., Ref.~\cite{tsintzis2022creating}. We do not consider the three remaining TRs in this work but note that these can also be connected to analytical solutions of the sweet spot condition. 

\section{\label{sec:including_Coulomb}Including on-site Coulomb repulsion}
The sweet spot condition given in Eq.~\eqref{eq:sweet_spot_condition} is only valid for $U=0$, where one can work with the simpler Hamiltonian given in Eq.~\eqref{eq:ham_tight_binding}. If $U \neq 0$, we must work with the second quantized Hamiltonian, given in Eq.~\eqref{eq:hamiltonian_second_quantization_explicit}, which makes the problem too complex to find sweet spots analytically. Thus, we have to use numerical calculations. Doing so, we could not find a sweet spot in any model studied with $U>0$. Instead, we find TRs with imperfect PMMs.
We will now consider the case when the effective couplings are transmitted via superconducting bulk states and the case when they are transmitted via ABSs separately.

\subsection{\label{sec:via_bulk_states}Transmission via superconducting bulk states}

\begin{figure*}
	\centering
	\includegraphics[width=\linewidth]{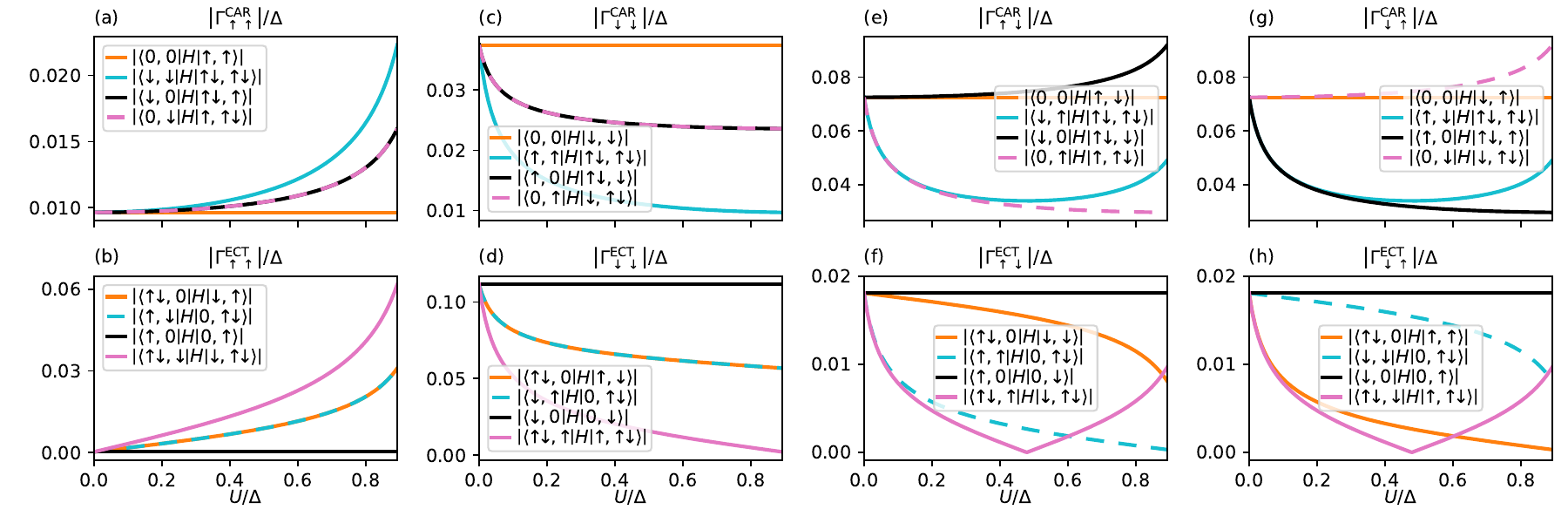}
	\caption{Impact of finite Coulomb repulsion on CAR [panels (a), (c), (e), and (g)] and ECT [panels (b), (d), (f), and (h)] that are transmitted via superconducting bulk states.
		All amplitudes are calculated to second order in $t/\Delta$.
		The analytical expressions are given in Appendix~ \ref{app:effective_theory}.
		If $U=0$, then there is only one distinct value for each ECT/CAR term.
		If $U > 0$, then there are three to four distinct values for each ECT/CAR term. This adds significant complexity to the model, rendering it impossible to study PMMs analytically. 
		The colors indicate the states that the corresponding amplitudes connect; see the Hamiltonian given in Eq.~\eqref{eq:hamiltonian_second_quantization_explicit}.
		The parameters are $\mu/\Delta=-0.479$, $\Delta_Z/\Delta=0.487$, $t/\Delta=0.1$, $\Phi_\mathrm{SOI} = 0.1\pi$. We note that these are the parameters where, in Fig.~\ref{fig:analytical_sweet_spot_bulk}, the black cross is.
	}
	\label{fig:effective_parameters_vs_u}
\end{figure*}

\begin{figure}
	\centering
	\includegraphics[width=\linewidth]{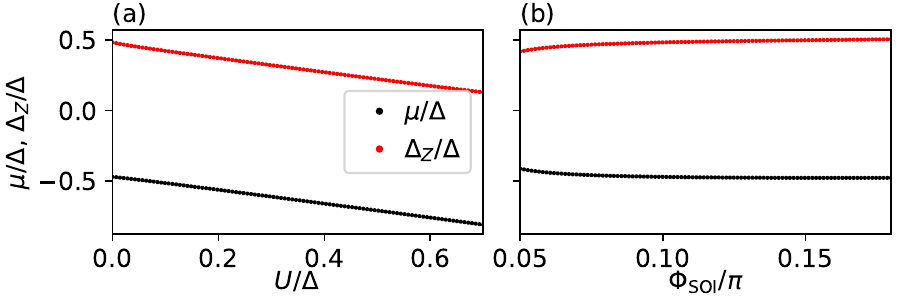}
	\caption{Dependencies of Zeeman energy $\Delta_Z$ and chemical potential $\mu$ that lead to a TR in systems where effective couplings are transmitted via superconducting bulk states. 
	In panel~(a), the on-site Coulomb interaction $U$ is increased and in panel~(b), the SOI angle $\Phi_\mathrm{SOI}$. The black dots indicate the corresponding value of the chemical potential $\mu$ and the red dots the value of the Zeeman energy $\Delta_Z$ at which a TR is found. 
	As discussed in Sec.~\ref{sec:reaching_sweet_spot_effective_models}, the TR at $U=0$ and $\Phi_\mathrm{SOI}=0.1\pi$ is connected to the sweet spot condition given in Eq.~\eqref{eq:sweet_spot_condition} and, as shown here, it is continuously connected to the TRs for finite $U>0$ and different values for the SOI angle. The parameters are the same as for Fig.~\ref{fig:analytical_sweet_spot_bulk}. In panel~(b), $U=0$.
	All TRs are within the threshold values $\Delta E_\mathrm{th}/\Delta = 10^{-6}$, $\Delta Q_\mathrm{th} = 0.03$, $M_\mathrm{th}=0.03$, and have a finite excitation gap $E_\mathrm{ex}$.		
	}
	\label{fig:rot_continuous_connection_bulk}
\end{figure}

The effective Hamiltonian is calculated using the equations given in Appendix~\ref{app:effective_theory}. In contrast to Sec.~\ref{sec:reaching_sweet_spot_effective_models}, as a consequence of the nonzero on-site Coulomb repulsion $U>0$, the CAR and ECT amplitudes depend on the occupation number of the QDs, i.e., different values for $\Gamma^\mathrm{ECT/CAR}_{\sigma\sigma',j}$ for different values of $j$; see Fig.~\ref{fig:effective_parameters_vs_u}. This results in a much more complicated model compared to Sec.~\ref{sec:sweet_spot}.

We do not find any sweet spots numerically. However, we find several TRs with a high MP. We can continuously connect these TRs to the TR found at $U=0$, which is related to the solution of the sweet spot condition given in Eq.~\eqref{eq:sweet_spot_condition}; see Fig.~\ref{fig:rot_continuous_connection_bulk}.
We describe the method for finding the TR in Appendix~\ref{app:finding_ROT_with_sigmoids}.

\subsection{\label{sec:via_ABSs}Transmission via ABSs}

\begin{figure}
	\centering
	\includegraphics[width=\linewidth]{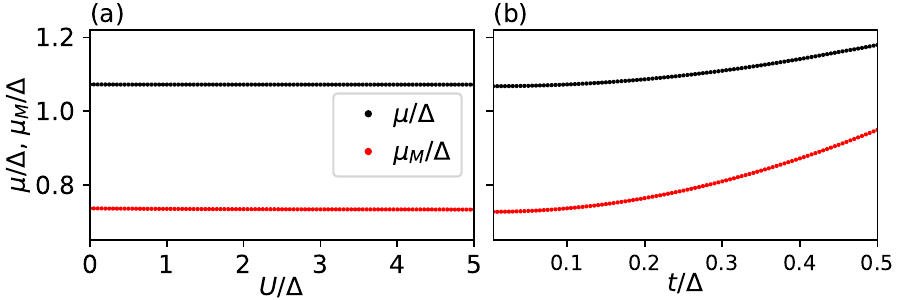}
	\caption{Dependencies of ABS and QD chemical potentials that lead to a TR in systems where effective couplings are transmitted via an ABS.
	In panel~(a), the Coulomb interaction $U$ increases and in panel~(b), the tunneling amplitude $t$ increases. The black (red) dots indicate the chemical potential $\mu$ ($\mu_M$) of the outer QDs (central QD), where the TR is found.
	As discussed in Sec.~\ref{sec:reaching_sweet_spot_effective_models}, the TR at $t/\Delta=0.1$ and $U=0$ is related to the sweet spot condition given in Eq.~\eqref{eq:sweet_spot_condition} and as shown here, this TR is continuously connected to the TRs at finite Coulomb interaction $U>0$ and larger hopping amplitudes $t$.
	The parameters for this figure are the same as for Fig.~\ref{fig:sweet_spot_analytics_ABS}. All TRs are within the threshold values $\Delta E_\mathrm{th}/\Delta = 10^{-6}$, $\Delta Q_\mathrm{th} = 0.02$, and $M_\mathrm{th}=0.02$.
	}
	\label{fig:rot_connection}
\end{figure}

We use the Hamiltonian given in Eq.~\eqref{eq:hamiltonian_ABS_simple} to search for a TR with finite on-site Coulomb repulsion $U$ and larger hopping amplitudes $t/\Delta$, see Appendix~\ref{app:finding_ROT_with_sigmoids} for more detail on the search for TRs. We do not find any sweet spots, however, there are several TRs with imperfect PMMs. These TRs are continuously connected to the TR found in Sec.~\ref{sec:reaching_sweet_spot_effective_models}, where $U=0$ and $t/\Delta \ll 1$, which were related to the sweet spot condition given in Eq.~\eqref{eq:sweet_spot_condition}; see Fig.~\ref{fig:rot_connection}.

\section{\label{sec:conclusion}Conclusion}
We have derived an analytical sweet spot condition for achieving perfectly localized zero-energy PMMs, i.e., perfect PMMs, in a spinful two QD system that has effective CAR, ECT, and LAR, but no Coulomb interaction terms. Reaching this sweet spot, however, requires independent control over all parameters in the model. We consider two more realistic models for transmitting CAR, ECT, and LAR: via superconducting bulk states, or via an ABS. We show that in both cases, the sweet spot condition can only be satisfied approximately, resulting in highly, but not perfectly, localized near-zero-energy states, called imperfect PMMs. We also find such states for finite Coulomb interaction and show that they are continuously connected to the sweet spot condition. However, using these more microscopic theories, no perfect PMMs are found due to parameter interdependencies. The definition of the imperfect PMMs in these more realistic models relies on threshold values, thus adding some arbitrariness to the concept of PMMs.

The more realistic models used in this work are still simplified compared to reality. Although we considered some dependencies between parameters, there are still relations we neglected. One such relation is the hopping amplitude between the QDs and the superconductor, which depends on the overlap of the wave functions. Thus, the hopping amplitude is not an independent parameter of the model. A more microscopic model that takes these dependencies into account is considered in Ref.~\cite{liu2024coupling}.
We have included the effect of the on-site Coulomb interaction $U$ on the model parameters via Schrieffer-Wolff perturbation theory. Electron-electron interactions, however, affect proximitized superconducting gaps to a greater extent, which was shown in Ref.~\cite{thakurathi2018Majorana}.
For the model described in Eq.~\eqref{eq:hamiltonian_ABS_simple}, we assumed that only a single ABS transmits ECT and CAR, whereas experiments suggest that several ABSs are involved in the process~\cite{dvir2023realization, bordin2023crossed, zatelli2023robust, bordin2024signatures}. Given that the dependencies considered in this paper already precluded perfect PMMs, it would be interesting to consider these additional interdependencies and study their effect on the quality of the imperfect PMMs.

\begin{acknowledgments}
This work was supported as a part of NCCR SPIN, a National Centre of Competence in Research, funded by the Swiss National Science Foundation (grant number 225153).
This project received funding from the European Union’s Horizon 2020 research and innovation program (ERC Starting Grant, Grant Agreement No. 757725).  H.F.L. acknowledges support by the Georg H. Endress Foundation. 
\end{acknowledgments}

\appendix

\section{\label{app:asymmetric}Threshold region in an asymmetric system}
\begin{figure}
		\centering
		\includegraphics[width=\linewidth]{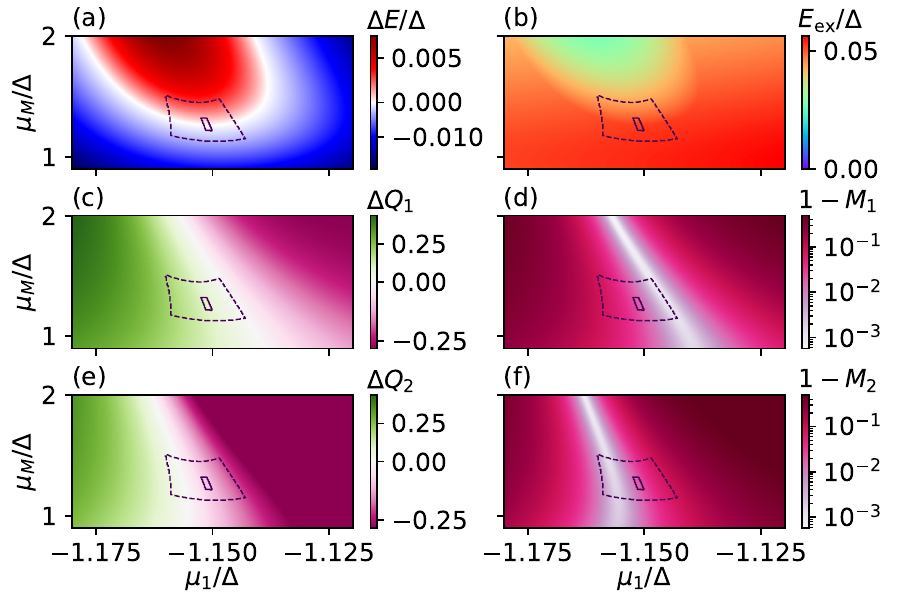}
		\caption{
				TRs for a model in which CAR and ECT are transmitted via an ABS, setup as shown in Fig.~\ref{fig:setup_ABS}. We vary the chemical potentials $\mu_1$ and $\mu_2 = 0.8 \mu_1$ of the two outer QDs and the chemical potential $\mu_M$ of the central QD to search for a TR. 
				In panel~(a), the energy difference $\Delta E$ between the even and odd ground states is shown, the excitation gap $E_\mathrm{ex}$ is shown in panel~(b), panel~(c) shows the charge difference $\Delta Q_j$ and the MP $M_j$ is shown in panel~(d). Because the system is not symmetric, we obtain $\Delta Q_1 \neq \Delta Q_2$ and $M_1 \neq M_2$. 
				We show two TRs with different threshold values: a dashed line with loose threshold values and a solid line for a TR with stricter threshold values.
				The parameters for this data are $\Delta_{Z,2} = 0.8 \Delta_{Z,1}$, $\Delta_{Z,1}/\Delta = -1.2$, $\Phi_\mathrm{SOI} = 0.15 \pi$, $t/\Delta= 0.3$, and $U=0$. 
				The threshold values for the dashed TR are $\Delta E_\mathrm{th}/\Delta = 3 \times 10^{-3}$,  $\Delta Q_\mathrm{th} = 0.2$, and $M_\mathrm{th} = 0.2$ and the threshold values for the solid TR are 
				$\Delta E_\mathrm{th}/\Delta = 10^{-3}$,  $\Delta Q_\mathrm{th} = 0.1$, and $M_\mathrm{th} = 0.1$. The excitation gap $E_\mathrm{ex}$ is finite in the TRs. 
		}
		\label{fig:rot_asymmetric}
	\end{figure}

	In this Appendix, we present data that demonstrates that a TR can exist even in a system that is not symmetric, i.e., the chemical potential and Zeeman energy on the QDs differ. We consider only the model in which effective couplings are transmitted via an ABS and replace the term $\sum_{j, \sigma} (\mu + \sigma \Delta_Z) d_{j\sigma}^\dagger d_{j\sigma}$ in Eq.~\eqref{eq:hamiltonian_ABS_simple} by the term $\sum_{j, \sigma} (\mu_j + \sigma \Delta_{Z,j}) d_{j\sigma}^\dagger d_{j\sigma}$. We set $\mu_2 = 0.8 \mu_1$, $\Delta_{Z,2} = 0.8 \Delta_{Z,1}$, resulting in a TR; see Fig.~\ref{fig:rot_asymmetric}. We note that the PMMs in these asymmetric cases generally tend to require looser threshold conditions to obtain similar areas in the TR compared to the symmetric case. It should be emphasised that this asymmetric case will generically be the case, since the fabrication of identical dots is a significant experimental challenge.

\section{\label{app:effective_theory}Derivation of the effective theory for transmission via superconducting bulk states}

In this Appendix, we derive the effective theory for the two cases in which the coupling between the QDs is either transmitted via superconducting bulk states or via an ABS. Since the derivations of the effective theory for these two cases are quite similar, we perform it side-by-side. We use Schrieffer-Wolff perturbation theory~\cite{winkler2003spin, bravyi2011schrieffer, scherubl2019transport, spethmann2023highfidelity} to integrate out the superconductor. The small parameter in which we expand the effective Hamiltonian is $t/\Delta$. In addition, we assume that the SOI vector is perpendicular to the magnetic field, i.e., $\mathbf{n}=(0,1,0)$ in Eq.~\eqref{eq:soi_matrix}.

In the case where the effective coupling is transmitted via superconducting bulk states, the setup is given by Fig.~\ref{fig:setup_3} and the corresponding Hamiltonian is given by Eq.~\eqref{eq:hamiltonian_sc_bulk_full}. For the Schrieffer-Wolff perturbation theory we define the unperturbed Hamiltonian $H_{\mathrm{bulk},0}$ as
\begin{equation}
H_{\mathrm{bulk},0} = H_\mathrm{bulk} (t=0),
\end{equation}
where $H_\mathrm{bulk}(t)$ is the Hamiltonian defined in Eq.~\eqref{eq:hamiltonian_sc_bulk_full}. For the superconducting part, we do a Bogoliubov transformation
\begin{subequations}
	\begin{align}
		H_{\mathrm{bulk,SC}} \! =& 
		\sum_{\mathbf{k}, \sigma} \! \epsilon_\mathbf{k} c_{\mathbf{k} \sigma}^\dagger c_{\mathbf{k}\sigma} 
		\! - \! \sum_{\mathbf{k}} \!
		\Delta  ( 
		e^{-i\phi} c_{\mathbf{k} \up}^\dagger c_{-\mathbf{k} \dn}^\dagger
		\! + \! e^{i\phi}
		c_{-\mathbf{k} \dn}
		c_{\mathbf{k} \up} 
		) 
		\nonumber \\  
		=& \sum_{\mathbf{k}\sigma} E(\epsilon_\mathbf{k}) \gamma_{\mathbf{k}\sigma}^\dagger \gamma_{\mathbf{k}\sigma}, \\
		E(\epsilon_\mathbf{k}) =& \sqrt{\Delta^2 + \epsilon_\mathbf{k}^2}, \\
		c_{\mathbf{k}\sigma} =& u(\epsilon_\mathbf{k}) \gamma_{\mathbf{k}\sigma} + \sigma e^{-i\phi} v(\epsilon_\mathbf{k}) \gamma_{-\mathbf{k} -\sigma}^\dagger, \\
		u(\epsilon_\mathbf{k}) =& \sqrt{
			\frac{1}{2} \left(
			1 + \frac{\epsilon_\mathbf{k}}{E(\epsilon_\mathbf{k})}
			\right),
		} \\
		v(\epsilon_\mathbf{k}) =& \sqrt{
			\frac{1}{2} \left(
			1 - \frac{\epsilon_\mathbf{k}}{E(\epsilon_\mathbf{k})}
			\right).
		} 
	\end{align}
\end{subequations}
The perturbation Hamiltonian is then given by
\begin{align}
H_{\mathrm{bulk},T} (t) = H_\mathrm{bulk}(t) - H_{\mathrm{bulk},0}.
\end{align}

In the case where transmission happens via an ABS, the setup is given by Fig.~\ref{fig:setup_ABS} and the Hamiltonian is defined in Eq.~\eqref{eq:hamiltonian_ABS_simple}. The unperturbed Hamiltonian $H_{\mathrm{ABS},0}$ is given by
\begin{equation}
H_{\mathrm{ABS},0} = H_{\mathrm{ABS}} (t=0),
\end{equation}
where $H_{\mathrm{ABS}} (t) $ is given by Eq.~\eqref{eq:hamiltonian_ABS_simple}. Once again, we do a Bogoliubov transformation on the superconducting part
\begin{subequations}
	\begin{align}
		H_\mathrm{SC,ABS} =&
		\sum_{\sigma} \mu_M c_\sigma^\dagger c_\sigma
		- 
		\Delta  ( 
		e^{-i\phi} c_\up^\dagger c_\dn^\dagger
		+ e^{i\phi}
		c_\dn
		c_\up
		) 
		\nonumber \\ 
		=& \sum_{\sigma} E(\mu_M) \gamma_\sigma^\dagger \gamma_\sigma, \\
		E(\mu_M) =& \sqrt{\Delta^2 + \mu_M^2}, \\
		c_\sigma =& u(\mu_M) \gamma_\sigma + \sigma e^{-i\phi} v(\mu_M) \gamma_{-\sigma}^\dagger,\\
		u(\mu_M) =& \sqrt{
			\frac{1}{2} \left(
			1 + \frac{\mu_M}{E(\mu_M)}
			\right)	}, \\
		v(\mu_M) =& \sqrt{
			\frac{1}{2} \left(
			1 - \frac{\mu_M}{E(\mu_M)}
			\right).} 
	\end{align}
\end{subequations}
The perturbation Hamiltonian is given by
\begin{align}
H_{\mathrm{ABS},T} (t) = H_{\mathrm{ABS}} (t) - H_{\mathrm{ABS},0}.
\end{align}

We thus see that the only difference between the case where the couplings are transmitted via superconducting bulk states and where they are transmitted via an ABS is that in the former, we sum over states with momentum $\mathbf{k}$ and energy $\epsilon_\mathbf{k}$, whereas in the latter case, there is no summation and there is only one state at $\mu_M$.

Using the same basis as introduced in Sec.~\ref{sec:setup_and_model} and doing a second-order Schrieffer-Wolff transformation in $t/\Delta$ gives the effective Hamiltonian $H_{\mathrm{eff},\mathrm{even}}$ and $H_{\mathrm{eff},\mathrm{odd}}$. This effective Hamiltonian has the same form as the Hamiltonian given in Eq.~\eqref{eq:hamiltonian_second_quantization_explicit}, but the parameters are no longer independent variables. 

Before we present the effective parameters, we illustrate how to transform a summation over $\mathbf{k}$ into an integral expression.  Let us assume that we consider a term $f_\mathbf{k} \equiv f(\Delta, \epsilon_\mathbf{k})$:
\begin{align} \label{eq:sum_k_to_integral_epsilon}
	\sum_\mathbf{k} f_\mathbf{k} 
	&=  \sum_\mathbf{k}  f(\Delta, \epsilon_\mathbf{k})
	 =  \intop_{-\infty}^\infty  d\epsilon \, \rho(\epsilon) f(\Delta, \epsilon) 
	\nonumber \\ &
	\approx \rho_F  \intop_{-\infty}^\infty  d\epsilon \, f(\Delta, \epsilon),
\end{align}
where $\rho(\epsilon)$ is the density of states, which is assumed to be constant and equal to $\rho_F$ in the last step. 
Since all calculations done in this work are limited to second-order perturbation theory, in any physically relevant quantity, the density of states $\rho_F$ appears only in the combination $\rho_F t^2$. Furthermore, the combination of $\rho_F \Delta$ is a dimensionless number, which we label $\rho_F \Delta = \kappa$. Any appearance of $\rho_F$ can therefore be written as $\rho_F t^2 = \kappa t^2/\Delta = (\sqrt{\kappa}t)^2 / \Delta = \tilde{\kappa} \tilde{t}^2$, where we have defined the rescaled parameters $\tilde{t} = \sqrt{\kappa} t$ and $\tilde{\kappa} = 1/\Delta$, corresponding to a rescaled density of states $\tilde{\rho}_F = 1/\Delta$. Therefore, the value of $\rho_F$ is irrelevant, since it can be absorbed in a rescaling of the hopping amplitude $t$. We emphasize that the value of the hopping amplitude $t$ given in the captions of Figs.~\ref{fig:analytical_sweet_spot_bulk},~\ref{fig:effective_parameters_vs_u}, and~\ref{fig:rot_continuous_connection_bulk} are, in fact, the effective values of the rescaled amplitude $\tilde{t} = \sqrt{\kappa} t$. 

In the following, we will label results that depend on the way ECT and CAR are transmitted by ``bulk'' (``ABS'') if the result is for the case where the effective coupling is transmitted via superconducting bulk states (via an ABS). Ignoring a constant shift in energy, the chemical potential, Zeeman energy, and on-site Coulomb interaction are renormalized:
\begin{subequations}
\begin{align}
\mu &\rightarrow \mu_\mathrm{eff} = \mu + \frac{B_+ + B_-}{2} - A, \\
\Delta_Z &\rightarrow \Delta_{Z,\mathrm{eff}} =  \Delta_Z + \frac{B_+ - B_-}{2}, \\
U &\rightarrow U_\mathrm{eff} = U + A + C - (B_+ + B_-), \label{eq:effective_u}
\end{align}
\end{subequations}
with
\begin{widetext}
\begin{subequations} \allowdisplaybreaks
	\begin{align}
		& A = -t^2 g_A^\mathrm{bulk/ABS}, \\
		& B_\pm =  -t^2 g_{B_\pm}^\mathrm{bulk/ABS}, \\
		& C = -t^2 g_C^\mathrm{bulk/ABS}, \\
& g_A^\mathrm{bulk}  =  \rho_F 
\intop_{-\infty}^\infty 
d \epsilon \, v^2 (\epsilon)
 \left( 
\frac{1}{E(\epsilon)-\Delta_Z+\mu}
 +  \frac{1}{E(\epsilon)+\Delta_Z+\mu}
 \right) , \\
& g_{B_\pm}^\mathrm{bulk}  =  \rho_F 
\intop_{-\infty}^\infty 
d\epsilon \left( 
\frac{u^2(\epsilon)}{E(\epsilon) \mp \Delta_Z - \mu}
 +  \frac{v^2(\epsilon)}{E(\epsilon)+U \mp \Delta_Z+\mu}
 \right)  , \\
& g_C^\mathrm{bulk} = \rho_F 
\intop_{-\infty}^\infty 
d \epsilon \, u^2(\epsilon)   \left( 
\frac{1}{E(\epsilon) - U - \Delta_Z - \mu}
 +   \frac{1}{E(\epsilon) - U + \Delta_Z - \mu}
 \right)  ,  \\
& g_A^\mathrm{ABS}  =  v^2(\mu_M) 
 \left( 
\frac{1}{E(\mu_M)-\Delta_Z+\mu}
 + \frac{1}{E(\mu_M)+\Delta_Z+\mu}
 \right)  , \\
& g_{B_\pm}^\mathrm{ABS} = 
\frac{u^2(\mu_M)}{E(\mu_M) \mp \Delta_Z - \mu}
+ \frac{v^2(\mu_M)}{E(\mu_M)+U \mp \Delta_Z+\mu},
 \\
& g_C^\mathrm{ABS} = u^2( \mu_M  )  \left( 
\frac{1}{E(  \mu_M  ) - U - \Delta_Z - \mu}
 +  \frac{1}{E(  \mu_M  ) - U + \Delta_Z - \mu}
 \right)  .
\end{align}
\end{subequations}
We note that in the case $U=0$, one obtains $A+C=B_+ + B_-$ and therefore $U_\mathrm{eff}=0$.

The terms coming from LAR processes are given by
{\allowdisplaybreaks
\begin{align}  
&\Gamma^\mathrm{LAR} =
	- t^2 e^{-i\phi} g^\mathrm{LAR, bulk/ABS}, \\
&g^\mathrm{LAR,bulk} =  \rho_F \!\!\!  \intop_{-\infty}^\infty  \!\!
d \epsilon \, \frac{u(\epsilon) v(\epsilon)}{2} \Bigg( \!
\frac{1}{E(\epsilon)-\Delta_Z+\mu}
	\!+\! \frac{1}{E(\epsilon)+\Delta_Z+\mu}
	\!+\! \frac{1}{E(\epsilon)-U-\Delta_Z-\mu}
	\!+\! \frac{1}{E(\epsilon)-U+\Delta_Z-\mu}
	 \! \Bigg), \\
&g^\mathrm{LAR,ABS} = \frac{u(\mu_M) v(\mu_M)}{2} \Bigg( \!
	\frac{1}{E(\mu_M)  - \Delta_Z + \mu}
	\! + \! \frac{1}{E(\mu_M) + \Delta_Z + \mu}
\! + \! \frac{1}{E(\mu_M) - U - \Delta_Z - \mu}
\! + \! \frac{1}{E(\mu_M) - U + \Delta_Z - \mu}
	\! \Bigg).
\end{align} }
We emphasize that $A$, $B_\pm$, $C$, and the LAR expressions do not depend on the SOI angle $\Phi_\mathrm{SOI}$.

The effective CAR couplings for two particles of the same spin are given by
\begin{subequations} \label{eq:CAR_samespin} \allowdisplaybreaks
	\begin{align} 
	&\Gamma^\mathrm{CAR}_{\sigma\sigma,j} = -
		t^2 e^{-i\phi} \sin(\Phi_\mathrm{SOI}) 
		\sin(\theta) g^\mathrm{CAR,bulk/ABS}_{\sigma\sigma,j},  j=1,2,\\
	&\Gamma^\mathrm{CAR}_{\sigma\sigma,3} = \Gamma^\mathrm{CAR}_{\sigma\sigma,4} = \frac{\Gamma^\mathrm{CAR}_{\sigma\sigma,1}+\Gamma^\mathrm{CAR}_{\sigma\sigma,2}}{2}, \\
&g^\mathrm{CAR,bulk}_{\sigma\sigma,1}  =  
\rho_F \intop_{-\infty}^\infty  d \epsilon \,
u(\epsilon) v(\epsilon) 
\left( 
\frac{1}{E(\epsilon) -\sigma\Delta_Z-\mu}
+ \frac{1}{E(\epsilon)+\sigma\Delta_Z+\mu}
 \right), \\
&g^\mathrm{CAR,bulk}_{\sigma\sigma,2} = \rho_F  \intop_{-\infty}^\infty  d \epsilon \,
		u(\epsilon) v(\epsilon) 
		\left(
		\frac{1}{E(\epsilon)-U-\sigma\Delta_Z-\mu}
		+ \frac{1}{E(\epsilon)+U+\sigma\Delta_Z+\mu}
		\right), \\
&g^\mathrm{CAR,ABS}_{\sigma\sigma,1} = 
 u(\mu_M) v(\mu_M) 
\left(
\frac{1}{E(\mu_M)-\sigma\Delta_Z-\mu}
+ \frac{1}{E(\mu_M)+\sigma\Delta_Z+\mu}
\right), \\
&g^\mathrm{CAR,ABS}_{\sigma\sigma,2} = 
u(\mu_M) v(\mu_M) 
\left(
\frac{1}{E(\mu_M)-U-\sigma\Delta_Z-\mu}
+ \frac{1}{E(\mu_M)+U+\sigma\Delta_Z+\mu}
\right).
\end{align}
\end{subequations}

The effective CAR coupling for particles of different spins is given by
\begin{subequations} \label{eq:CAR_diffspin}
\begin{align}
&\Gamma^\mathrm{CAR}_{\updn,1} = -\Gamma^\mathrm{CAR}_{\dnup,1} \frac{\lambda^*}{\lambda} = \frac{\Gamma^\mathrm{CAR}_{\upup,1}+\Gamma^\mathrm{CAR}_{\dndn,1}}{2 \lambda} ,\\
&\Gamma^\mathrm{CAR}_{\updn,2} = -\Gamma^\mathrm{CAR}_{\dnup,2} \frac{\lambda^*}{\lambda}  = \frac{\Gamma^\mathrm{CAR}_{\upup,2}+\Gamma^\mathrm{CAR}_{\dndn,2}}{2 \lambda}, \\
&\Gamma^\mathrm{CAR}_{\updn,3} = -\Gamma^\mathrm{CAR}_{\dnup,4} \frac{\lambda^*}{\lambda}  = \frac{\Gamma^\mathrm{CAR}_{\upup,2}+\Gamma^\mathrm{CAR}_{\dndn,1}}{2 \lambda},\\
&\Gamma^\mathrm{CAR}_{\updn,4} = -\Gamma^\mathrm{CAR}_{\dnup,3} \frac{\lambda^*}{\lambda}  = \frac{\Gamma^\mathrm{CAR}_{\upup,1}+\Gamma^\mathrm{CAR}_{\dndn,2}}{2 \lambda} ,
\end{align}
\end{subequations}
with
\begin{align} \label{eq:lambda}
\lambda = \frac{\sin (\theta)}{\frac{1}{\tan(\Phi_\mathrm{SOI})}-i\cos(\theta)},
\end{align}
which simplifies to $\tan(\Phi_\mathrm{SOI})$ if $\theta=\pi/2$, i.e., if the SOI direction is perpendicular to the magnetic field. We note that the expressions given in Eqs.~\eqref{eq:CAR_diffspin} and~\eqref{eq:lambda} are only defined if SOI is neither zero nor parallel to the magnetic field.

{\allowdisplaybreaks
The effective ECT couplings for particles of the same spin are given by
\begin{subequations}
\begin{align} 
&\Gamma^\mathrm{ECT}_{\sigma\sigma,1} = \Gamma^\mathrm{ECT}_{\sigma\sigma,2} =  \frac{\Gamma^\mathrm{ECT}_{\sigma\sigma,3}+\Gamma^\mathrm{ECT}_{\sigma\sigma,4}}{2}, \\
&\Gamma^\mathrm{ECT}_{\sigma\sigma,j}  = 
t^2  [ 
\cos (\Phi_\mathrm{SOI}) 
 -  i \sigma  \cos( \theta ) \sin( \Phi_\mathrm{SOI} ) 
] g^\mathrm{ECT,bulk/ABS}_{\sigma\sigma,j} ,  
\nonumber \\ & \quad
j=3,4,\\
&g^\mathrm{ECT,bulk}_{\sigma\sigma,3}  = 
\rho_F  \intop_{-\infty}^\infty  d \epsilon 
		 \left(
		-\frac{u^2(\epsilon)}{E(\epsilon) - \sigma\Delta_Z - \mu}
		 +  \frac{v^2(\epsilon)}{E(\epsilon) + \sigma\Delta_Z + \mu}
		 \right) , \\
&g^\mathrm{ECT,bulk}_{\sigma\sigma,4}  =  
		\rho_F  \intop_{-\infty}^\infty   d \epsilon
		 \left(
		-\frac{u^2(\epsilon)}{E(\epsilon) - U - \sigma\Delta_Z - \mu}
		+ \frac{v^2(\epsilon)}{E(\epsilon) + U + \sigma\Delta_Z + \mu}
		 \right) , \\
&g^\mathrm{ECT,ABS}_{\sigma\sigma,3}   = 
		-\frac{u^2(\mu_M)}{E(\mu_M)-\sigma\Delta_Z-\mu}
		+ \frac{v^2(\mu_M)}{E(\mu_M)+\sigma\Delta_Z+\mu}, \\
&g^\mathrm{ECT,ABS}_{\sigma\sigma,4} = 
		-\frac{u^2(\mu_M)}{E(\mu_M)-U-\sigma\Delta_Z-\mu}
		+ \frac{v^2(\mu_M)}{E(\mu_M)+U+\sigma\Delta_Z+\mu}.
	\end{align}
\end{subequations}
}
\end{widetext}

The effective ECT couplings for particles of different spins are given by
\begin{subequations}
\begin{align}
& \Gamma^\mathrm{ECT}_{\updn,1}= -(\Gamma^\mathrm{ECT}_{\dnup,2})^* = -\frac{\Gamma^\mathrm{ECT}_{\upup,4} \lambda +\Gamma^\mathrm{ECT}_{\dndn,3} \lambda^*}{2}, \\
& \Gamma^\mathrm{ECT}_{\updn,2} = -(\Gamma^\mathrm{ECT}_{\dnup,1})^* = -\frac{\Gamma^\mathrm{ECT}_{\upup,3}\lambda +\Gamma^\mathrm{ECT}_{\dndn,4}\lambda^* }{2}, \\
& \Gamma^\mathrm{ECT}_{\updn,3} = -(\Gamma^\mathrm{ECT}_{\dnup,3})^* = -\frac{\Gamma^\mathrm{ECT}_{\upup,3} \lambda+\Gamma^\mathrm{ECT}_{\dndn,3} \lambda^*}{2}, \\
& \Gamma^\mathrm{ECT}_{\updn,4} = -(\Gamma^\mathrm{ECT}_{\dnup,4})^* = -\frac{\Gamma^\mathrm{ECT}_{\upup,4}\lambda +\Gamma^\mathrm{ECT}_{\dndn,4} \lambda^*}{2}.
\end{align}
\end{subequations}
We note that none of the ECT expressions depend on the superconducting phase $\phi$.

We emphasize that the quantities $g_{A/B_\pm/C}^\mathrm{bulk/ABS}$, $g^\mathrm{LAR,bulk/ABS}$, and $g_{\sigma\sigma'}^\mathrm{CAR/ECT,bulk/ABS}$ are only introduced to simplify the notation, but these are not physically relevant quantities. Therefore, the statement made above, that $\rho_F$ only appears in the combination $\rho_F t^2$ in physically relevant quantities still holds.

We emphasize that in the case where ECT and CAR are transmitted via superconducting bulk states, there are quite stringent conditions on the parameters to assure convergence of the integrals. These conditions are
\begin{align} \label{eq:condition_parameters_integral_bulk}
\left|
\mu + U + \left|\Delta_Z\right|
\right| < \Delta 
\text{ and }
\left|\mu - \left|\Delta_Z\right|\right| < \Delta.
\end{align}
These conditions guarantee that the states on the QDs are inside the bulk superconductor gap.
It does not hold if ECT and CAR are transmitted via an ABS.

If $U=0$, then $\Gamma^\mathrm{CAR/ECT}_{\sigma\sigma',j} \equiv \Gamma^\mathrm{CAR/ECT}_{\sigma\sigma'}$ (and $g_{\sigma\sigma',j}^\mathrm{CAR/ECT,bulk/ABS} = g_{\sigma\sigma'}^\mathrm{CAR/ECT,bulk/ABS}$) for all $j=1, \dots, 4$. In this case, and setting $\phi=0$, the ECT and CAR matrices have the form
\begin{widetext}
\begin{subequations}
\begin{align}
\mathcal{H}_\mathrm{T,ECT} &= C t^2
\begin{pmatrix}
e^{-i\rho} g_\upup^\mathrm{ECT,bulk/ABS} 
& - \frac{e^{-i\rho} \lambda g_\upup^\mathrm{ECT,bulk/ABS} + e^{i\rho} \lambda^\ast g_\dndn^\mathrm{ECT,bulk/ABS} }{2} \\[4pt]
\frac{e^{i\rho} \lambda^\ast  g_\upup^\mathrm{ECT,bulk/ABS} + e^{-i\rho} \lambda g_\dndn^\mathrm{ECT,bulk/ABS} }{2} 
& e^{i\rho} g_\dndn^\mathrm{ECT,bulk/ABS} 
\end{pmatrix}, \\[5pt]
\mathcal{H}_\mathrm{T,CAR} &= -t^2 K
\begin{pmatrix}
	  g_\upup^\mathrm{CAR,bulk/ABS}
	& \frac{g_\upup^\mathrm{CAR,bulk/ABS} + g_\dndn^\mathrm{CAR,bulk/ABS}}{2\lambda} \\[4pt]
	- \frac{g_\upup^\mathrm{CAR,bulk/ABS} + g_\dndn^\mathrm{CAR,bulk/ABS}}{2\lambda^\ast}
	&  g_\dndn^\mathrm{CAR,bulk/ABS}
\end{pmatrix},
\end{align}
\end{subequations}
where $Ce^{i\rho}$ and $K$ are defined in Eq.~\eqref{eq:definition_C_rho_K}. To simplify further, we use the explicit expression of $\lambda$, $Ce^{i\rho}$, and $K$ to find $Ce^{-i\rho} \lambda = K$, thus giving:
\begin{subequations}
	\begin{align}
		\mathcal{H}_\mathrm{T,ECT} &= t^2
		\begin{pmatrix}
			Ce^{-i\rho} g_\upup^\mathrm{ECT,bulk/ABS} 
			& - K \frac{g_\upup^\mathrm{ECT,bulk/ABS} + g_\dndn^\mathrm{ECT,bulk/ABS} }{2} \\[4pt]
			K \frac{ g_\upup^\mathrm{ECT,bulk/ABS} +  g_\dndn^\mathrm{ECT,bulk/ABS} }{2} 
			& Ce^{i\rho} g_\dndn^\mathrm{ECT,bulk/ABS} 
		\end{pmatrix}, \\[5pt]
		\mathcal{H}_\mathrm{T,CAR} &= -t^2 
		\begin{pmatrix}
			K g_\upup^\mathrm{CAR,bulk/ABS}
			& Ce^{-i\rho} \frac{g_\upup^\mathrm{CAR,bulk/ABS} + g_\dndn^\mathrm{CAR,bulk/ABS}}{2} \\[4pt]
			- C e^{i\rho} \frac{g_\upup^\mathrm{CAR,bulk/ABS} + g_\dndn^\mathrm{CAR,bulk/ABS}}{2}
			& K g_\dndn^\mathrm{CAR,bulk/ABS}
		\end{pmatrix},
	\end{align}
\end{subequations}
\end{widetext}
which has, up to overall complex conjugation, the same form as the ansatz we made in Eq.~\eqref{eq:t_and_delta_explicit}.

\section{\label{app:gap}Excitation gap at the sweet spot}
\begin{figure}
	\centering
	\includegraphics[width=\linewidth]{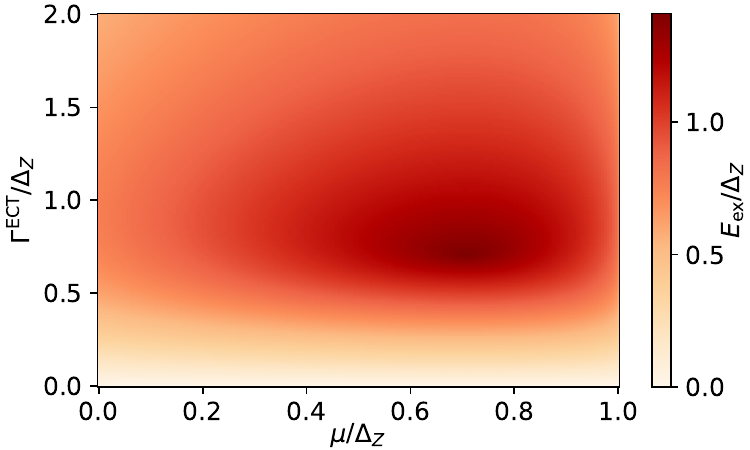}
	\caption{
		Excitation gap $E_\mathrm{ex}$, as defined in Eq.~\eqref{eq:definition_gap}, at the sweet spot using the same parameters as for Fig.~\ref{fig:deviation_various_parameters_1}. Ideally, one strives to maximize the excitation gap to achieve a more stable configuration.}
	\label{fig:gap_at_sweet_spot}
\end{figure}

In this Appendix, we consider the simplified model defined in the caption of Fig.~\ref{fig:deviation_various_parameters_1} at the sweet spot. The excitation gap $E_\mathrm{ex}$, defined in Eq.~\eqref{eq:definition_gap}, varies in parameter space; see Fig.~\ref{fig:gap_at_sweet_spot}. Thus, although the excitation gap is finite, there is an ideal point in parameter space leading to a maximum in $E_\mathrm{ex}$.

\section{\label{app:sweet_spot_analytics}Approximating the sweet spot condition}
In this Appendix, we give more details on how we find an approximate solution to the sweet spot condition given in Eq.~\eqref{eq:sweet_spot_condition}. We emphasize that we set $U=0$.

\subsection{Transmission via superconducting bulk states}
To find the best approximate solution of Eq.~\eqref{eq:sweet_spot_condition} for Fig.~\ref{fig:analytical_sweet_spot_bulk}, we use the numerically calculated effective parameters given in Appendix~\ref{app:effective_theory} and use SciPy's~\cite{2020SciPy-NMeth} basinhopping method to minimize the function
\begin{widetext}
\begin{align}
f(\mu, \Delta_Z) = &
\frac{1}{\Delta_Z^2} \Big[ 
\big|
\mu_\mathrm{eff}^2 - \Delta_{Z,\mathrm{eff}}^2 + (\Gamma^\mathrm{LAR})^2
\big|
 +  \big| 
(\mu_\mathrm{eff}  +  \Delta_{Z,\mathrm{eff}})
(\Gamma^\mathrm{CAR}_\updn  -  s \Gamma^\mathrm{ECT}_\updn )
 -  \Gamma^\mathrm{LAR} (
\Gamma^\mathrm{ECT}_\upup  -  s\Gamma^\mathrm{CAR}_\upup
)
\big|
\nonumber \\ &
 +  \big|
(\mu_\mathrm{eff}  +  \Delta_{Z,\mathrm{eff}})
 (\Gamma^\mathrm{CAR}_\dndn   -  s \Gamma^\mathrm{ECT}_\dndn )
 -  (\mu_\mathrm{eff}  -  \Delta_{Z,\mathrm{eff}})
 (\Gamma^\mathrm{CAR}_\upup  -  s\Gamma^\mathrm{ECT}_\upup )
\big| 
\Big],
\end{align}
which is the sum of the absolute values of the differences between the left-hand sides and the right-hand sides of Eq.~\eqref{eq:sweet_spot_condition}, divided by $\Delta_Z^2$ to make the expression dimensionless and to numerically discourage solutions with small magnetic fields. In the optimization process, we exclude any solution that does not satisfy the parameter limits given in Eq.~\eqref{eq:condition_parameters_integral_bulk}.

\subsection{Transmission via ABSs}
If ECT and CAR are transmitted via ABSs, the expressions for the effective couplings to second order in $t$ do not contain any integrals and therefore, we can do a more analytical approach compared to the case where transmission happens via superconducting bulk states.

Since $U=0$, one obtains that all $\Gamma^\mathrm{CAR/ECT}_{\sigma\sigma',j}$ for $j=1,2,3,4$ are equal. We will therefore label them as $\Gamma^\mathrm{CAR/ECT}_{\sigma\sigma'}$. Since we only calculate the effective parameters up to second order in $t/\Delta$, we will also consider the sweet spot conditions only up to second order in $t/\Delta$. Additionally, we set $\theta=0$. Therefore, Eq.~\eqref{eq:sweet_spot_condition} becomes
\begin{subequations} \label{eq:sweet_spot_condition_second_order}
\begin{align}
0 =& \mu_\mathrm{eff}^2 - \Delta_{Z,\mathrm{eff}}^2 + (\Gamma^\mathrm{LAR})^2
= \mu^2 - \Delta_Z^2 + t^2 \big[
2 \mu g_A^\mathrm{ABS}
- (\mu-\Delta_Z) g_{B_+}^\mathrm{ABS}
- (\mu+\Delta_Z) g_{B_-}^\mathrm{ABS}
\big] , \\
0 
=& (\mu_\mathrm{eff} + \Delta_{Z,\mathrm{eff}} )
(s \Gamma^\mathrm{CAR}_\updn - \Gamma^\mathrm{ECT}_\updn )
- \Gamma^\mathrm{LAR} (s \Gamma^\mathrm{ECT}_\upup-\Gamma^\mathrm{CAR}_\upup )
\nonumber \\
=& \frac{t^2}{2} (\mu + \Delta_Z ) \big[
(
g^\mathrm{ECT,ABS}_\upup + g^\mathrm{ECT,ABS}_\dndn
) \sin \Phi_\mathrm{SOI}
- s (
g^\mathrm{CAR,ABS}_\upup + g^\mathrm{CAR,ABS}_\dndn
) \cos \Phi_\mathrm{SOI}
\big] , \\
0 
=& (\mu_\mathrm{eff} + \Delta_{Z,\mathrm{eff}}) (s \Gamma^\mathrm{CAR}_\dndn -  \Gamma^\mathrm{ECT}_\dndn)
- (\mu_\mathrm{eff} - \Delta_{Z,\mathrm{eff}}) (s \Gamma^\mathrm{CAR}_\upup - \Gamma^\mathrm{ECT}_\upup) 
\nonumber \\
=& - t^2  \big[ 
(\mu+\Delta_Z) (g^\mathrm{ECT,ABS}_\dndn \cos \Phi_\mathrm{SOI} + s g^\mathrm{CAR,ABS}_\dndn \sin \Phi_\mathrm{SOI} )
 -  (\mu-\Delta_Z) (g^\mathrm{ECT,ABS}_\upup \cos \Phi_\mathrm{SOI} + s g^\mathrm{CAR,ABS}_\upup \sin \Phi_\mathrm{SOI} ) 
\big].
\end{align}
\end{subequations}
We can simplify these equations further, and, assuming that $\mu+\Delta_Z \neq 0$, we obtain
\begin{subequations}
\begin{align} 
0 =& \mu^2 - \Delta_Z^2 + t^2 \left[
\frac{(\Delta_Z-\mu)(\Delta_Z+\mu+\mu_M)}{\Delta^2+\mu_M^2 - (\Delta_Z+\mu)^2}
+ \frac{(\Delta_Z+\mu)(\Delta_Z-\mu-\mu_M)}{\Delta^2+\mu_M^2 - (\Delta_Z-\mu)^2}
\right] , \\
0 =& \Delta (
- \Delta^2 + \Delta_Z^2 + \mu^2 - \mu_M^2
) \cos \Phi_\mathrm{SOI}
+ s \big[
-\Delta^2 (\mu+\mu_M)
+ (\mu-\mu_M)
 (
-\Delta_Z^2 + [\mu+\mu_M]^2
)
\big] \sin \Phi_\mathrm{SOI} , \\
0 =&  [
\Delta^2 \mu_M
+ (\Delta_Z-\mu-\mu_M)
(2\mu-\mu_M)
(\Delta_Z+\mu+\mu_M)
] \cos \Phi_\mathrm{SOI} 
- s \Delta (
\Delta^2 - \Delta_Z^2 - 3 \mu^2 + \mu_M^2
) \sin \Phi_\mathrm{SOI}.
\end{align}
\end{subequations}
\end{widetext}
If we fix $t/\Delta$ and $\Phi_\mathrm{SOI}$, then we can numerically solve this system of equations for $\mu/\Delta$, $\mu_M/\Delta$, and $\Delta_Z/\Delta$.
We stress that this is only a solution to second order in $t$. The full model, as defined by Eq.~\eqref{eq:hamiltonian_ABS_simple} does not satisfy the sweet spot condition exactly and therefore, there are no perfect PMMs in this system.

\section{\label{app:finding_ROT_with_sigmoids} Finding a threshold region numerically}
To find a TR numerically, one can either sweep through the parameters and identify all points that satisfy Eq.~\eqref{eq:definition_ROT}. If, however, one has to do this for many parameters, e.g., in Fig.~\ref{fig:rot_connection}, this brute-force approach is very time-consuming. Therefore, we use SciPy's~\cite{2020SciPy-NMeth} basinhopping function to find a global minimum of the function $f(\mathbf{p})$:
\begin{align}
f(\mathbf{p})
=& 
\sigma \left(|\Delta Q_1(\mathbf{p})| + |\Delta Q_2(\mathbf{p})| ; a_Q, x_{0,Q}\right) 
\nonumber \\ &
+ \sigma \left(2-M_1(\mathbf{p})-M_2(\mathbf{p}) ; a_M, x_{0,M}\right) ,
\nonumber \\ & 
\text{ with } \Delta E (\mathbf{p}) = 0 ,\\
\sigma(x; a, x_0) =& [1+e^{-a\left(x-x_0\right)}]^{-1} 
\end{align}
where $\mathbf{p}$ indicates all parameters that are varied, e.g., $\mathbf{p}=(\mu, \mu_M)$ for Fig.~\ref{fig:sweet_spot_analytics_ABS} or $\mathbf{p}=(\mu, \Delta_Z)$ for Fig.~\ref{fig:analytical_sweet_spot_bulk}, $a_Q$, $a_M$, $x_{0,Q}$, and $x_{0,M}$ are parameters that are chosen such that values for $\Delta Q_j$ and $1-M_j$ below the TR threshold give a small contribution to $f(\mathbf{p})$, while values above the TR threshold give a large contribution to $f(\mathbf{p})$. We use $a_Q=49$, $a_M=461$, $x_{0,Q}=0.16$, and $x_{0,M}=0.035$. The condition $\Delta E(\mathbf{p})=0$ can be set as a constraint for the optimization process, i.e., only points in parameter space  that give $\Delta E(\mathbf{p})=0$ are considered as valid solutions for the optimization.
Our results show that the optimized point $\mathbf{p}$ lies within or just next to the TR.

\bibliography{bibliography}

\end{document}